\documentclass[preprint*]{JHEP3} % 10pt is ignored!

\JHEPspecialurl{http://jhep.sissa.it/JOURNAL/JHEP3.tar.gz}

\usepackage{epsfig,multicol}
\usepackage{amsmath,amssymb,bm}

\newcommand\fverb{\setbox\pippobox=\hbox\bgroup\verb}
\newcommand\fverbdo{\egroup\medskip\noindent%
                        \fbox{\unhbox\pippobox}\ }
\newcommand\fverbit{\egroup\item[\fbox{\unhbox\pippobox}]}
\newbox\pippobox
\title{
Solving the local cohomology problem in U(1) chiral
gauge theories within a finite lattice
}

\author{
Daisuke Kadoh, Yoshio Kikukawa and Yoichi Nakayama\\
Department of Physics, Nagoya University, Nagoya 464-8602, Japan\\
E-mail: \email{kadoh@eken.phys.nagoya-u.ac.jp}, 
        \email{kikukawa@eken.phys.nagoya-u.ac.jp}}
\preprint{\heplat{0309022}}      % OR: \preprint{Aaaa/Mm/Yy\\Aaa-aa/Nnnnnn}
\abstract{
In the gauge-invariant construction of abelian chiral gauge 
theories on the lattice based on %the Dirac operator satisfying 
the Ginsparg-Wilson relation, the gauge anomaly is topological 
and its cohomologically trivial part  plays the role of the local 
counter term. We give a prescription to solve the local cohomology 
problem within a finite lattice by reformulating the Poincar\'e lemma 
so that it holds true on the finite lattice up to exponentially 
small corrections. We then argue that the path-integral measure 
of Weyl fermions can be constructed directly from the quantities 
defined on the finite lattice.
}


\begin{document} 

%\maketitle  IS IGNORED %%%%%%%%%%%

\section{Introduction}
\label{sec:intro} 

Recently it turned out that lattice gauge theory can provide
a framework for non-perturbative study of  chiral gauge theories, despite 
the well-known problem of the species doubling. The clue
to this development is 
the construction of gauge-covariant and local lattice Dirac operators 
satisfying the Ginsparg-Wilson relation\cite{Ginsparg:1981bj,
Neuberger:1997fp,Hasenfratz:1998ri,Neuberger:1998wv,
Hasenfratz:1998jp,Hernandez:1998et},
\begin{equation}
\gamma_5 D + D \gamma_5  = 2 a D \gamma_5 D .
\end{equation}
An explicit solution of 
the Ginsparg-Wilson relation was derived from 
the overlap formalism based on domain wall fermion and is 
referred as the overlap Dirac operator.\footnote{
The overlap formalism proposed by Narayanan and 
Neuberger\cite{
Narayanan:wx,Narayanan:sk,Narayanan:ss,Narayanan:1994gw,Narayanan:1993gq,
Neuberger:1999ry,Narayanan:1996cu,Huet:1996pw,
Narayanan:1997by,Kikukawa:1997qh} 
gives a well-defined partition function of Weyl fermions on the lattice,
which nicely reproduces the fermion zero mode and the fermion-number
violating observables
('t Hooft vertices)\cite{Narayanan:1996kz,Kikukawa:1997md,Kikukawa:1997dv}. 
Through the recent re-discovery of the Ginsparg-Wilson relation,
the meaning of the overlap formula, especially
the locality properties, become clear from the point of view 
of the path-integral. 
The gauge-invariant construction by L\"uscher\cite{Luscher:1998du} 
based on the Ginsparg-Wilson relation 
provides a procedure to determine the phase of the overlap formula
 in a gauge-invariant manner for anomaly-free chiral gauge theories.
For Dirac fermions, the overlap formalism provides a
gauge-covariant and local lattice Dirac operator
satisfying the Ginsparg-Wilson relation\cite{Ginsparg:1981bj,
Neuberger:1997fp,Kikukawa:1997qh,Neuberger:1998wv,
Hernandez:1998et}. The overlap formula was 
derived from the five-dimensional approach of 
domain wall fermion proposed by Kaplan\cite{Kaplan:1992bt}.
In its vector-like formalism\cite{Shamir:1993zy, Furman:ky,
Blum:1996jf, Blum:1997mz}, 
the local low energy effective action of the chiral mode 
precisely reproduces the overlap Dirac 
operator \cite{Vranas:1997da,Neuberger:1997bg, Kikukawa:1999sy}.
}
It has made possible
to realize exact chiral symmetry on the
lattice\cite{Luscher:1998pq} and also opened a
route to the gauge invariant construction of anomaly-free chiral gauge 
theories on the lattice\cite{Luscher:1998kn,Luscher:1998du,Luscher:1999un,
Luscher:1999mt,Luscher:2000hn}. 

In the gauge-invariant construction of chiral gauge theories on the lattice, 
one of the crucial steps is 
to establish the exact cancellation of gauge anomaly at a finite
lattice spacing. In abelian chiral gauge 
theories\cite{Luscher:1998du,Suzuki:1999qw}\footnote{
See also \cite{Neuberger:2000wq}
for a gauge-invariant construction of abelian 
chiral gauge theories in non-compact formulation.
},
it has been achieved through the cohomological classification of the chiral 
anomaly\cite{Luscher:1998kn,Fujiwara:1999fi,Fujiwara:1999fj}, 
which is given in terms of 
lattice Dirac operator satisfying the Ginsparg-Wilson 
relation\cite{Luscher:1998pq,Kikukawa:1998pd, 
Fujikawa:1998if, Adams:1998eg, Suzuki:1998yz, Chiu:1998xf},
\begin{equation}
\label{eq:chiral-anomaly}
q(x) = \text{tr}\left\{ \gamma_5(1-a D)(x,x)\right\}
\end{equation}
and is a topological field 
for the admissible lattice gauge fields satisfying the 
bound\footnote{
It has been shown by Neuberger \cite{Neuberger:1999pz} that
the constant $1/30$ in the above bounds can be improved to
$1/6(2+\sqrt{2})$.
}
\begin{eqnarray}
\label{eq:admissibility-1}
&& \parallel 1-P_{\mu\nu}(x)
\parallel <  \epsilon, \quad \epsilon  < \frac{1}{30} \\
\label{eq:admissibility-2}
&& P_{\mu\nu}(x) \equiv U(x,\mu)U(x+\hat \mu,\nu)U(x+\hat
\nu,\mu)^{-1}U(x,\nu)^{-1}. 
\end{eqnarray}
For an anomaly-free multiplet of Weyl fermions satisfying
the anomaly cancellation condition of the U(1) charges,
\begin{equation}
\sum_\alpha e_\alpha^3 = 0,
\end{equation}
it has been shown that the chiral anomaly is cohomologically trivial,
\begin{equation}
\label{eq:cohomological-triviality-of-chiral-anomaly}
\sum_\alpha e_\alpha q^\alpha(x) = \partial_\mu^\ast k_\mu(x), 
\quad\quad
q^\alpha(x) = \left. q(x) \right\vert_{U\rightarrow U^{e_\alpha}} ,
\end{equation}
where $k_\mu(x)$ is a certain gauge-invariant and local current. 
The cohomologically trivial part of the chiral 
anomaly is then used in the gauge-invariant  construction of the 
Weyl fermion measure. In short, it plays the role of the 
local counter term in the effective action for the Weyl 
fermions.\footnote{
For nonabelian chiral gauge theories, the local
cohomology problem can be formulated with 
the topological field in 4+2 dimensional
space.\cite{Alvarez-Gaume:1983cs,Luscher:1999un,
Luscher:1999mt,
Luscher:2000hn,Adams:2000yi} So far, the exact cancellation of 
gauge anomaly has been shown in all orders of the perturbation
expansion for generic nonabelian theories\cite{Suzuki:2000ii, Igarashi:2000zi,
Luscher:2000zd},
and nonperturbatively for $SU(2)\times U(1)_Y$ electroweak 
theory, both in the infinite lattice\cite{Kikukawa:2000kd}. 
In the five-dimensional approach using the 
domain wall fermion\cite{Kaplan:1992bt,
Shamir:1993zy, Furman:ky,
Blum:1996jf, Blum:1997mz, 
Neuberger:1997bg, Kikukawa:1999sy}, the local cohomology problem
can be formulated in 5+1 dimensional space\cite{Kikukawa:2001mw}. 
}

For the practical computation of
observables in the lattice abelian chiral gauge theories,
it is required to compute the Weyl fermion measure for 
every admissible configuration. 
However it seems difficult
to follow the steps given in \cite{Luscher:1998du} literally.
The first problem is the use of the infinite lattice in order
to make sure the locality property of  
the cohomologically trivial part.\footnote{
The cohomologically trivial part is, so far,
constructed in two steps:
the local cohomology problem is first solved in the
infinite lattice\cite{Luscher:1999mt} and then the corrections required
in the finite lattice are constructed and 
added\cite{Igarashi:2002zz} . 
Since the lattice Dirac operator satisfying the 
Ginsparg-Wilson relation should have the exponentially
decaying tail\cite{Horvath:1998cm,Horvath:1999bk}, 
the local fields in consideration should have the infinite 
number of components. Moreover, the vector potentials used in 
this analysis are not bounded.
}
As a closely related problem,
the vector-potential-representative of the link variable used in the
cohomological analysis is unbounded.
The second problem is the use of the continuous interpolations
in the space of the admissible $U(1)$ gauge fields.

The purpose of this paper is to give a prescription
to solve the local cohomology problem {\it within a finite lattice}
which applies directly to the chiral anomaly on the finite lattice,
\begin{equation}
q(x)={\rm tr}\left\{
\gamma_5(1- D_L)(x,x) \right\},  
\end{equation}
where $D_L(x,y)$ is the finite-volume kernel of the Dirac operator. 
With this method, 
we will show that 
the current $k_\mu(x)$, which gives the 
the cohomologically trivial part of the above chiral anomaly,
can be obtained  directly from the quantities calculable on the finite lattice. 
Then we will argue that 
the measure of the Weyl fermions can be also constructed
directly from the quantities defined on the finite 
lattice.\footnote{
The continuous interpolation
in the space of the admissible gauge fields 
is also a crucial technique in the above construction.
We will discuss the issue
how to implement the continuous interpolation numerically
in the forthcoming paper.
}
For our purpose, we first examine the Poincar\'e lemma,  which is 
originally formulated on the infinite lattice\cite{Luscher:1998kn}. 
We will show that the lemma can be reformulated
so that it holds true on the
finite lattice up to exponentially small corrections of
order
$O({\rm e}^{-L/2\varrho})$, where $L$ is the 
lattice size and $\varrho$ is the 
localization range of the differential forms in 
consideration.\footnote{
For the ultra-local tensor fields on a finite lattice,
the Poincar\'e lemma has been formulated by Fujiwara et al in 
\cite{Fujiwara:2000wn}.
} (Section~\ref{sec:poincare-lemma-on-finite-lattice})

We next examine the one-to-one correspondence between the link variable
and the vector potential with a certain good
locality property, which is originally
derived in the infinite lattice\cite{Luscher:1998kn}.
We will show that a similar one-to-one correspondence 
with the desired locality property can be formulated for 
the admissible 
U(1) gauge fields on the finite lattice,
by separating the link variables into the part which 
is responsible to the magnetic flux (the constant mode
of the field strength) and the part of the local and dynamical 
degrees of freedom around the magnetic flux. 
(Section~\ref{sec:vector-potential-on-finiteV})

Equipped with 
the modified Poincar\'e lemma and the vector potentials for 
the admissible U(1) gauge fields on the finite lattice, 
we will perform the cohomological analysis of the 
chiral anomaly directly on the finite lattice. 
Through this analysis,
we will derive a formula by which 
the cohomologically trivial part of 
the chiral anomaly is given
in terms of the quantities defined on the finite lattice. 
(Section~\ref{sec:chiral-anomaly-in-finite-lattice})

In this paper, we will focus on how to 
obtain the cohomologically trivial part (the local counter term)
directly on the finite lattice.
We do not claim that the cohomological classification of 
the chiral anomaly can be completed within the setup of
the finite lattice. 
Rather we will use some of the results in the infinite
volume as inputs in order to establish the exact cancellation of
gauge anomaly on the finite lattice, 
as we will see in section~\ref{sec:chiral-anomaly-in-finite-lattice}.
For the presentation of our result, 
we follow closely the convention and the notation adopted 
in \cite{Luscher:1998kn}, so that the necessary modification
of the cohomological analysis in the finite lattice 
becomes clear.

\section{U(1) gauge fields on the finite lattice}
\label{sec:gauge-fields-on-finiteV}

We consider a finite four-dimensional lattice of size $L$ with periodic
boundary conditions and choose lattice units.
The U(1) gauge fields on such a lattice may be represented through periodic
link fields, 
\begin{eqnarray}
U(x,\mu) \in {\rm U(1)}, \qquad\qquad x=(x_1,\cdots,x_4) \in
\mathbb{Z}^4,
\\
U(x+L \hat \nu,\mu)=U(x,\mu) \ \  {\rm for \ all}  \ \
\mu,\nu=1,\cdots,4,\hspace{0.1mm}
\end{eqnarray}
on the infinite lattice. The independent degrees of 
freedom are then the link variables at the point $x$
in the block
\begin{equation}
\Gamma_{4\,[x_0]} =
 \left\{x \in \mathbb{Z}^4 \vert -L/2 \le x_\mu-x_{0\mu} < L/2 \right\}
\end{equation}
where $x_0$ is a certain reference point. 
$L$ is assumed to be an even integer. 
Under gauge transformations
\begin{equation}
U(x,\mu) \rightarrow \Lambda(x) U(x,\mu) \Lambda(x+\hat\mu)^{-1},
\end{equation}
the periodicity of the field will be preserved if 
$\Lambda(x) \in $ U(1) is periodic.

We impose the so-called admissibility condition on the U(1)
gauge fields:
\begin{equation}
\vert F_{\mu\nu}(x) \vert  < \epsilon \ \  {\rm for \ all} \ \ 
x,\mu,\nu, 
\end{equation}
where the field tensor $F_{\mu\nu}(x)$ is defined through
\begin{eqnarray}
F_{\mu\nu}(x) &=& \frac{1}{i} {\rm ln} P_{\mu\nu}(x), 
\quad - \pi < F_{\mu\nu}(x) \le \pi .
\end{eqnarray}
We require this condition because it ensures that 
the overlap Dirac 
operator\cite{Neuberger:1997fp,Neuberger:1998wv} is a
smooth and local function of   the gauge field for $\epsilon <
1/30$ \cite{Hernandez:1998et}.

The admissible U(1) gauge fields on the finite lattice
can be classified by the magnetic fluxes $m_{\mu\nu}$, 
where
\begin{equation}
m_{\mu\nu} 
=
\frac{1}{2\pi}\sum_{s,t=0}^{L-1}F_{\mu\nu}(x+s\hat\mu+t\hat\nu)
\end{equation}
(integers independent of $x$). 
The following field is 
periodic and can be shown
to have constant field tensor equal to 
$2\pi m_{\mu\nu}/L^2 (< \epsilon)$:
\begin{eqnarray}
V_{[m]}(x,\mu) 
&=&{\text{e}}^{-\frac{2\pi i}{L^2}\left[
L \delta_{\tilde x_\mu,L-1} \sum_{\nu > \mu} m_{\mu\nu}
\tilde x_\nu +\sum_{\nu < \mu} m_{\mu\nu} \tilde x_\nu
\right]}, %\nonumber\\
\end{eqnarray}
where the abbreviation $\tilde x_\mu = x_\mu$ mod $L$
has been used. Then any admissible U(1) gauge field
in the topological sector with the magnetic flux $m_{\mu\nu}$
may be expressed as 
\begin{equation}
U(x,\mu)=\tilde U(x,\mu) \, V_{[m]}(x,\mu) .
\end{equation}
We may regard $\tilde U(x,\mu)$ as 
the actual local and dynamical degrees of freedom
in the given topological sector.

\section{Local composite fields on the finite-volume lattice}
\label{sec:chiral-anomaly-in-finite-lattice-and-locality}

Under the admissibility condition,  the space of the U(1) lattice gauge 
fields on the finite-volume lattice is separated into topological sectors labeled by the 
magnetic flux $m_{\mu\nu}$. Over the topological sectors, the gauge
field can not be deformed continuously to each other without breaking the 
admissibility condition. Then a composite field of gauge field,
which is a functional defined over the field space, could be
defined independently in each topological sectors. 
Moreover, even in each topological sectors, the link variables are not 
independent each other and then the locality of composite
fields of gauge field is not completely obvious. Therefore we need to clarify 
what exactly a local composite field on the finite-volume lattice means.

We first note the fact  
that the finite-volume kernel 
$D_L(x,y)$ of a lattice Dirac operator of a Ginsparg-Wilson type fermion
may be represented by the kernel in the infinite lattice $D(x,y)$ 
which is restricted to the periodic link field: 
\begin{equation}
\label{eq:finite-volume-kernel}
  D_L(x,y)= D(x,y) + \sum_{n \in \mathbb{Z}^4,n \not = 0}
  D(x,y + n L), 
\end{equation}
\begin{equation}
D_L(x,y)= D(x,y) + {\rm O}( {\rm e}^{- L/\varrho} ) . 
\end{equation}
The second equation follows from the requirement of the 
locality for the lattice Dirac operator $D$ in the infinite 
lattice\cite{Hernandez:1998et, Luscher:1998kn}.  
Namely, it is required  
that $D(x,y)$ is a sum of strictly local operators,
\begin{equation}
D(x,y)= \sum_{k=1}^\infty D_k(x,y), 
\end{equation}
which are localized on the blocks with side-lengths $2k$. 
Moreover these kernels and their 
derivatives $D_k(x,y;y_1,\nu_1,\ldots,y_n,\nu_n)$ with respect to 
the gauge field variables $U(y_1,\nu_1)$, $\ldots$, $U(y_n,\nu_n)$
are required to satisfy the bounds
\begin{equation}
\label{eq:bound-on-k-kernel-on-infinite-lattice}
\| D_k(x,y;y_1,\nu_1,\ldots,y_n,\nu_n) \| \le a_n k^{p_n} {\rm e}^{- \theta k }  
\qquad (n \ge 0)
\end{equation}
where the constants $a_n, p_n \ge 0$ and $\theta > 0 $ can 
be chosen to be independent of the gauge field. 
As a consequence we have
\begin{equation}
\label{eq:bound-kernel-infinite-lattice}
\| D(x,y;y_1,\nu_1,\ldots,y_n,\nu_n) \| \le a_n^\prime
(1 + \| x -z \|^{p_n} ) {\rm e}^{- \|x-z\| / \varrho } ,
\end{equation}
for a constant $a_n^\prime$ and the integer $p_n \ge 0$, 
where $z$ is chosen from $y,y_1,\ldots,y_n$ so that $ \| x -z \|$ is the maximum. 
The localization range is given by $\varrho= 2/ \theta$. 
In the case of the overlap Dirac operator, it has been proved that
these requirements are satisfied for all admissible gauge 
fields\cite{Hernandez:1998et}.

From this observation, it is reasonable, for our purpose, to specify the 
notion of a local composite field on the finite-volume lattice as follows: 

\vspace{1em}
\noindent {\bf Locality property of  local composite fields (I)} : 
{\it we refer a 
composite field $\phi(x)$ on the finite periodic lattice  as local if 
$\phi(x)$ can be expressed as the sum of two parts, 
the local composite field $\phi_\infty(x)$ defined
in the infinite lattice and  restricted with a periodic gauge field
and the finite-volume correction $\Delta \phi_\infty(x)$:
\begin{equation}
\label{eq:phi-separated}
\phi(x)=\phi_\infty(x) + \Delta \phi_\infty(x) \, ;  \qquad
\vert \Delta \phi_\infty(x) \vert  \le  c \, L^{q} \, {\rm e}^{- L/ 2 \varrho} 
%          =\phi_\infty(x)+ {\rm O}( {\rm e}^{- L/\varrho} ) ,  
\end{equation}
for a constant $c$ and an integer $p \ge 0$, 
where $\varrho$ is the localization range of $\phi(x)$.  
$\phi_\infty(x)$ is required to be 
local in the sense that it can be written as a series
\begin{equation}
\label{eq:sum-in-ultra-local-fields}
\phi_\infty(x)= \sum_{k=1}^\infty \phi_{k\infty}(x) 
\end{equation}
 of strictly local fields $\phi_k(x)$  which are localized on the blocks
 with side-lengths proportional to $k$, 
 and these fields and their derivatives 
 $\phi_k (x;y_1,\nu_1,\ldots,y_m,\nu_m)$ with respect to
the gauge field variables are bounded by
\begin{equation}
\label{eq:locality-bound-for-ultra-local-fields}
\vert \phi_{k \infty}(x;y_1,\nu_1,\ldots,y_m,\nu_m)\vert \le 
c_m k^{q_m} {\rm e}^{- k /\varrho } \qquad (m \ge 1) , 
\end{equation}
where the constants $c_m, q_m \ge 0$ and the localization range $\varrho > 0 $ are 
independent of the gauge field. 
}

As to the locality of a composite field on the finite-volume
lattice, it turns out to be convenient, for our purpose, 
to specify its property in more detail. 
We note again eq.~(\ref{eq:finite-volume-kernel}), 
the relation of the finite-volume kernel 
$D_L(x,y)$ to that in the infinite lattice $D(x,y)$, and also the fact that
the differentiation of $D_L(x,y)$ with respect to 
the periodic link field $U(z,\nu)$ is related to  
the differentiation of $D(x,y)$ with respect to the generic link field in the 
infinite lattice as follows: let $\tilde \eta(x,\nu)$ be a periodic variation of the link field,
while $\eta(x,\nu)$ is an arbitrary variation in the infinite lattice and then we have
\begin{eqnarray}
\frac{\delta D_L(x,y)}{\delta \tilde \eta (z,\nu)}
&=& \frac{\delta D(x,y)}{\delta \tilde \eta (z,\nu)} +
\sum_{n \in \mathbb{Z}^4,n \not = 0}\frac{\delta D(x,y+ n L)}{\delta \tilde \eta(z,\nu)}
\nonumber\\
&=&
\frac{\delta D(x,y)}{\delta \eta (z,\nu)}
+\sum_{m \in \mathbb{Z}^4,m \not = 0}
\frac{\delta D(x,y)}{\delta  \eta (z+mL,\nu)} 
+ \sum_{n \in \mathbb{Z}^4,n \not = 0}\sum_{m \in \mathbb{Z}^4}
\frac{\delta D(x,y+ n L)}{\delta \eta(z+m L,\nu)}, \nonumber\\
\end{eqnarray}
where, after the variation, parameters $\tilde \eta$ and $\eta$ are set to be zero. 
From these relations we can see that
$D_L(x,y)$ is also a sum of operators,  
\begin{eqnarray}
\label{eq:finite-lattice-D-expansion-in-k}
D_L(x,y)&=& \sum_{k=1}^\infty D_{L k}(x,y),  \\
D_{L k} (x,y)&=& D_k(x,y) + \sum_{n \in \mathbb{Z}^4,n \not = 0}D_k(x,y + n L), 
\end{eqnarray}
where $D_{L k} (x,y)$ is strictly local in the sense that
the kernel itself and their derivatives 
$D_{L k}(x,y;y_1,\nu_1,\ldots,y_n,\nu_n)$ with respect to 
the periodic gauge field variables $U(y_1,\nu_1)$, $\ldots$, $U(y_n,\nu_n)$
vanishes identically if  
$2k  < \underset{z=y,y_1,\ldots,y_n} {max} \| x - z \|$:
\begin{equation}
D_{L k} (x,y;y_1,\nu_1,\ldots,y_n,\nu_n) = 0 \qquad
\text{if} \quad  2k < \underset{z=y,y_1,\ldots,y_n}{max} \| x - z \| . 
\end{equation}
Moreover we can infer 
the following bounds:
\begin{equation}
\| D_{L k}(x,y;y_1,\nu_1,\ldots,y_n,\nu_n) \| \le b_n k^{p_n} {\rm e}^{- \theta k } 
\qquad (n \ge 0) , 
\end{equation}
where the constants $b_n, p_n \ge 0$ and $\theta > 0 $ can 
be chosen to be independent of the gauge field and the lattice size $L$. 
As a consequence we have
\begin{equation}
\label{eq:bound-kernel-finite-lattice}
\| D_L (x,y;y_1,\nu_1,\ldots,y_n,\nu_n) \| \le b_n^\prime
(1 + \| x -z \|^{p_n} ) {\rm e}^{- \|x-z\| / \varrho }
\qquad (n \ge 0)  , 
\end{equation}
for a constant $b_n^\prime$ and the integer $p_n \ge 0$, 
where $z$ is chosen from $y,y_1,\ldots,y_n \in \Gamma_{4\,[x]}$ 
so that $ \| x -z \|$ is the maximum.\footnote{For the overlap Dirac operator in the 
finite-volume lattice, the expansion eq.~(\ref{eq:finite-lattice-D-expansion-in-k}) 
may be the Legendre polynomial expansion with respect to the operator $z$ 
which involves the square of the Wilson-Dirac operator in the finite-volume lattice,
\begin{equation}
  D_{w L}(x,y)= D_w(x,y) + \sum_{n \in \mathbb{Z}^4,n \not = 0}
  D_w(x,y + n L).
\end{equation}
With this expansion in the finite-volume lattice,  
the same argument as in 
\cite{Hernandez:1998et} 
holds true and   
the  locality bound eq.~(\ref{eq:bound-kernel-finite-lattice}) follows
for the lattice points $x$ and $y,y_1,\ldots,y_n \in  \Gamma_{4\,[x]}$. 
}
Therefore, it is reasonable to require further the following property
for a local composite field on the finite-volume lattice:

\vspace{1em}
\noindent {\bf Locality property of  local composite fields (II)} : 
{\it 
a local composite field $\phi(x)$ on the finite periodic lattice 
is also expressed as the sum of local fields, 
\begin{equation}
\label{eq:phi-expansion-in-k}
\phi(x)= \sum_{k=1}^\infty \phi_k(x) , 
\end{equation}
where $\phi_k(x)$ and their derivatives $ \phi_k (x;y_1,\nu_1,\ldots,y_m,\nu_m)$
with respect to 
the periodic gauge field variables $U(y_1,\nu_1)$, $\ldots$, $U(y_m,\nu_m)$
satisfy  
\begin{eqnarray}
\label{eq:phi-k-localization-range}
&& \phi_k (x;y_1,\nu_1,\ldots,y_m,\nu_m) = 0 \qquad \text{if} \quad
 2k < \underset{z=y,y_1,\ldots,y_n}{max} \| x - z \| , \\
\label{eq:phi-k-bound}
&& \vert \phi_k(x;y_1,\nu_1,\ldots,y_m,\nu_m)\vert \le 
d_m k^{q_m} {\rm e}^{- k /\varrho } \qquad (m \ge 1 )
\end{eqnarray}
for a constant $d_m$ and the integer $q_m \ge 0$. 
where the constants $d_m, q_m \ge 0$ and the localization range $\varrho > 0 $ are 
independent of the gauge field and  of the lattice size $L$. 
}

\vspace{1em}
\noindent 
Then it follows that
\begin{equation}
\label{eq:bound-phi-finite-lattice}
\| \phi(x;y_1,\nu_1,\ldots,y_m,\nu_m) \| \le d_m^\prime
(1 + \| x -z \|^{q_m} ) {\rm e}^{- \|x-z\| / \varrho }  \qquad (m \ge 1 )
\end{equation}
for a constant $d_m^\prime$, 
where $z$ is one of  $y_1,\ldots,y_m \in \Gamma_{4\,[x]}$ 
for which  $ \| x -z \|$ is the maximum.

Thus we adopt the notion of local composite fields on the finite 
periodic lattice defined by the properties (I) and (II).  
In the following sections, we will develop 
a method of the cohomological analysis  which directly applies to
this kind of local fields on the finite lattice.

\section{The Poincar\'e lemma on the finite lattice}
\label{sec:poincare-lemma-on-finite-lattice}

In this section, we reformulate the Poincar\'e lemma 
given in \cite{Luscher:1998kn} for the finite volume lattice.
We consider 
the finite lattice of integer vectors
$x=(x_1,\cdots,x_n) \in \mathbb{Z}^n$ in $n$ dimensions 
where $n \ge 1$ is left unspecified,
\begin{equation}
\Gamma_n = \left\{ x \in \mathbb{Z}^n \vert -L/2 \le x_\mu 
< L/2 \right\} .
\end{equation}
The size of the lattice, $L$, is assumed to be an 
even integer for simplicity. 

In the following we will be concerned with tensor fields 
$f_{\mu_1\cdots\mu_k}(x)$ on $\Gamma_n$ that are totally 
anti-symmetric in the indices $\mu_1,\cdots,\mu_k$ which
may be regarded as periodic tensor fields on the infinite 
lattice:
\begin{equation}
f_{\mu_1\cdots\mu_k}(x+L\hat\nu)=f_{\mu_1\cdots\mu_k}(x) \ \ 
{\rm for \ all} \ \ \mu,\nu=1,\cdots, n.
\end{equation}
The differential forms on the finite lattice are 
introduced following \cite{Luscher:1998kn}. 
The general $k$-form on $\Gamma_n$ is given by 
\begin{equation}
f(x) = \frac{1}{k!} f_{\mu_1\cdots\mu_k}(x) 
dx_{\mu_1}\cdots dx_{\mu_k}.
\end{equation}
The linear space of all these forms is denoted
by $\Omega_k$.
An exterior difference operator ${\rm d} : \Omega_k 
\rightarrow \Omega_{k+1}$ is defined through
\begin{equation}
\label{eq:def-k-form}
d f(x) = \frac{1}{k!} \partial_\mu 
f_{\mu_1,\cdots,\mu_k}(x) dx_\mu dx_{\mu_1}\cdots dx_{\mu_k},
\end{equation}
where $\partial_\mu$ denotes the forward nearest-neighbor
difference operator. 
The associated divergence operator $d^\ast : \Omega_k 
\rightarrow \Omega_{k-1}$ is defined in the obvious way
by setting $d^\ast f =0 $ if $f$ is a 0-form and
\begin{equation}
d^\ast f(x) = \frac{1}{(k-1)!} \partial_\mu^\ast
f_{\mu\mu_2\cdots\mu_k}(x) dx_{\mu_2}\cdots dx_{\mu_k}
\end{equation}
in all other cases, where $\partial_\mu^\ast$ is the backward
nearest-neighbor difference operator. 
With respect to the natural scalar product for tensor fields
on $\Gamma_n$, $d^\ast$ is equal to minus 
the adjoint operator of $d$. 

It is straightforward to show that ${d^\ast}^2=0$ and 
the difference equation $d^\ast f = 0$ is hence solved by all
forms $f = d^\ast g$. In the infinite lattice, 
the Poincar\'e lemma\cite{Luscher:1998kn} asserts
that these are in fact all solutions, 
an exception being the $0$-forms where one has a one-dimensional 
space of further solutions and one needs
the extra condition, $\sum_x f(x)=0$, to remove them.  
On the finite lattice, the lemma can be formulated
so that it holds true up to 
a certain correction form $\Delta f(x)$ which coefficients
are just some linear combinations of the coefficients of 
$f(x)$ at the boundary of $\Gamma_n$.\footnote{
An equivalent formulation of the Poincar\'e lemma can be given
in terms of divergence operator $d$.
}

\vspace{1em}
\noindent{\bf Lemma \ref{sec:poincare-lemma-on-finite-lattice}a \,(Modified
Poincar\'e lemma)} \\
Let $f$ be a $k$-form which satisfies
\begin{equation}
\label{eq:lemma-ast-condition}
d^\ast f = 0 \ \ \ {\rm and} \ \ \  \sum_{x\in \Gamma_n} f(x) =0 \ \ {\rm
if } \ \ k=0. 
\end{equation}
Then there exist two forms $g \in \Omega_{k+1}$ and 
$\Delta f \in \Omega_{k}$ such that
\begin{equation}
f = d^\ast g + \Delta f 
\end{equation}
for certain constants $C_3$ and $p_3 \ge 0$.
The tensor field $\Delta f_{\mu_1\cdots\mu_k}(x)$ linearly depends 
only on the values of the tensor field $f_{\mu_1\cdots\mu_k}(x)$ at the boundary, 
$\{ f_{\mu_1\cdots\mu_k}(z) \vert \ z \in  \partial \Gamma_n \}$. 

\vspace{1em}

On the other hand, 
in the continuum, it follows from the 
de Rham theorem that 
a closed form whose integral over any cycle vanishes identically
can be expressed as an exact form.
On the finite lattice $\Gamma_n$, 
any exact k-form($k \ge 1$),  $f=d^\ast g$,  satisfies 
\begin{equation}
\sum_{x_\nu;\nu \not ={\mu_1},\cdots,{\mu_k}}  f_{\mu_1\cdots\mu_k}(x)
=0  \qquad
\text{for all } {\mu_1},\cdots,{\mu_k}, 
\end{equation}
or, using $\partial^\ast_\mu f_{\mu\mu_2\cdots \mu_k}(x)=0$, 
\begin{equation}
\sum_{x \in \Gamma_n}  f_{\mu_1\cdots\mu_k}(x)
=0  \qquad
\text{for all } {\mu_1},\cdots,{\mu_k}. 
\end{equation}
Then we can show the following lemma. 

\vspace{1em}
\noindent{\bf Lemma \ref{sec:poincare-lemma-on-finite-lattice}b \, 
(Corollary of de Rham theorem)} \\
Let $f$ be a $k$-form which satisfies
\begin{equation}
\label{eq:lemma-ast-condition-b}
d^\ast f = 0 \ \  {\rm and} \ \  \sum_{x\in \Gamma_n} f(x) =0 .
\end{equation}
Then there exist a form $g \in \Omega_{k+1}$ 
such that
\begin{equation}
f = d^\ast g .
\end{equation}

\vspace{1em}
The proof of the  lemma 2a [lemma 2b] 
goes just like that in \cite{Luscher:1998kn} with
some modifications. 
We first
show that the lemma holds for $n=1$ and then proceed to 
higher dimensions by induction. On a one-dimensional 
lattice the non-trivial forms are the 0- and 1-forms. In 
the first case we have $k=0$ and the only condition on $f$ is
then that $\sum_{x \in \Gamma_1} f(x)=0$. It follows from 
this that the 1-form 
\begin{equation}
g(x) = \sum^{x_1}_{y_1={x_0}_1-L/2} f(y) \, \, d x_1
\end{equation}
is periodic and satisfies
$d^\ast g = f$. In the other case the equation $d^\ast f = 0$,
that is, $\partial_\mu^\ast f_\mu(x)=0$ implies that 
$f_1(x)=$ constant. 
Since one may choose $x$ as the point at the boundary 
$\| x-x_0 \| = L/2$,
we have $f_1(x) = f_1(z)\vert_{z \in \partial \Gamma_1}$ 
which is what the lemma claims. Thus we have 
proved the lemma for $n=1$.

\noindent
[If the one-form $f$ satisfies $\sum_{x_1 \in \Gamma_1} f_1(x_1)=0$, the
constant must vanishes identically and the second version of the 
lemma holds for $n=1$.]

Let us now assume that $n$ is greater than 1 and
that the lemma holds in dimension $n-1$. We then decompose
the form $f$ according to 
\begin{equation}
 f = u dx_n + v, 
\end{equation}
where $u$ and $v$ are elements of $\Omega_{k-1}$ and $\Omega_k$
respectively that are independent of $dx_n$.

If we ignore the
dependence on $x_n$, these forms may be regarded as forms
in $n-1$ dimensions. To avoid any confusion the corresponding
exterior divergence operator will be denoted by $\bar
d^\ast$.  It is then straightforward to show that 
\begin{equation}
d^\ast f = ( \bar d^\ast u ) \, dx_n + \{ (-1)^{k-1} \partial^\ast_n u  
+ \bar d^\ast v \} 
\end{equation}
and the equation $ d^\ast f = 0$ hence implies 
\begin{equation}
\label{eq:lemma-ast-condition-bar-u}
(-1)^{k-1} \partial^\ast_n u  
+ \bar d^\ast v =0, \qquad \bar d^\ast u =0.
\end{equation}
Note that $u=0$ if $k=0$ and the condition
eq.~(\ref{eq:lemma-ast-condition}) reduces to
$\sum_{x\in \Gamma_n} v(x)=0$. 

We now define a form $\bar v$ on $\Gamma_{n-1}$ through
\begin{equation}
\label{eq:lemma-ast-bar-v}
\bar v(x) = \sum_{y_n=-L/2}^{L/2-1} v(y), 
\qquad
y=(x_1,\cdots,x_{n-1}, y_n).
\end{equation}
Evidently $\bar v$ is an element of $\Omega_{k}$ and
from the above one infers that it satisfies the premises
of the lemma. The induction hypothesis thus allows us 
to conclude that $\bar v = \bar d^\ast \bar g + \Delta \bar v$ 
for some form $\bar g \in \Omega_{k+1}$ and 
some correction form $\Delta \bar v \in \Omega_{k}$
in $n-1$ dimensions. $\Delta \bar v$ depends only on 
$ \{ \bar v(z) \vert \  z \in \partial \Gamma_{n-1} \}$. 

Next we introduce a new form $h$ on $\Gamma_n$ through
\begin{equation}
h(x) = (-1)^{k} \sum_{y_n={x_0}_n-L/2}^{x_n}
\left\{ v(y) - \delta_{y_n,{x_0}_n} \bar v \right\} \, d x_n
\end{equation}
where $y$ is as in eq.~(\ref{eq:lemma-ast-bar-v}).
$h$ is periodic.
Using eq.~(\ref{eq:lemma-ast-condition-bar-u})
it is straightforward  to prove that 
\begin{eqnarray}
\label{eq:f-in-terms-h}
f(x)= \delta_{x_n,{x_0}_n} \bar v(x) + d^\ast h(x)
      + u(x)\vert_{x_n={x_0}_n-L/2-1}\, dx_n . 
\end{eqnarray}

$u(x)\vert_{x_n={x_0}_n+L/2-1}$  may be regarded as 
the $(k-1)$-form on $\Gamma_{n-1}$. To make it explicit, 
we denote the form as
\begin{equation}
u(x)\vert_{x_n={x_0}_n+L/2-1} = \bar u(x) .
\end{equation}
Since the second condition in eq.~(\ref{eq:lemma-ast-condition-bar-u})
implies that $\bar d^\ast \bar u = 0$, 
it follows 
\begin{equation}
\bar u(x) = 
\bar d^\ast \bar e(x)
+ \Delta \bar u(x), 
\end{equation}
where $\Delta \bar u$ depends only on 
$ \{ \bar u(z) \vert \  z \in \partial\Gamma_{n-1} \}$.
We thus conclude that the sum 
\begin{equation}
g(x) = \delta_{x_n,{x_0}_n} \bar g(x) + h(x)
+ \bar e(x) \, dx_n
\end{equation}
is an element of $\Omega_{k+1}$ such that
\begin{equation}
f(x) = d^\ast g(x) +  \Delta f(x) 
\end{equation}
where 
\begin{equation}
\Delta f(x) = \delta_{x_n,{x_0}_n} \Delta \bar v(x) 
+ \Delta \bar u(x) \, d x_n.
\end{equation}
$\Delta f $ depends only on 
$\{ \bar v(z) \vert \  z \in \Gamma_{n-1} \}$ and 
$ \{ \bar u(z) \vert \  z \in \Gamma_{n-1} \}$
and  
therefore depends only on the boundary values of $f(x)$, 
$ \{ f(z) \vert \  z \in \partial\Gamma_n\}$.

\noindent
[For the form $f$ satisfying $\sum_{x \in \Gamma_n} f(x)=0$, 
it follows from eq.~(\ref{eq:f-in-terms-h}) that 
$\sum_{x \in \Gamma_{n-1}} \bar v(x)=0$ 
and $\sum_{x \in \Gamma_{n-1}} u(x)\vert_{x_n={x_0}_n+L/2-1}
     =\sum_{x \in \Gamma_{n-1}} \bar u(x)=0$.
From the induction hypothesis, it follows immediately that 
$\Delta \bar v(x) = 0$ and 
$\Delta \bar u(x)=0$.
Therefore we have $f(x) = d^\ast g(x) $ and the second version
of the lemma holds in dimension $n$.]

The construction of the forms $g(x)$ and $\Delta f(x)$ is given explicitly in the above proof of the lemma. 
The coefficients of $g(x)$ and $\Delta f(x)$ are some linear combinations of the 
coefficients of $f(x)$ and therefore the sizes of these forms are intimately related to 
that of $f(x)$. 
Now let us introduce norms of the forms by 
\begin{eqnarray}
\|f \|_{x_0,p,\varrho} & = &\underset{\mu_1\cdots\mu_k,x \in \Gamma_n}{\rm max}\,
 \frac{|f_{\mu_1\cdots\mu_k}(x+x_0)|}{(1+\| x\|^{p} ) \,   {\rm e}^{-\| x\|/\varrho} },
\end{eqnarray}
with  a localization range $\varrho$,  an integer $p$ and 
a reference point $x_0$ fixed. 
Then we can show the following bound for the norm of the form $g(x)$:
\begin{equation}
\label{eq:bound-f-g}
\|g \|_{x_0,p,\varrho} \le C \, \|f \|_{x_0,p,\varrho}
\end{equation}
for some constant $C$ independent of $f(x)$, $x_0$ and $L$. 
The proof of this bound is given in the 
appendix~\ref{app:bound-on-norm-of-forms}.  As for the form 
$\Delta f(x)$, we have at least 
\begin{equation}
\label{eq:bound-delta-f}
\vert \Delta f_{\mu_1\cdots\mu_{k}}(x) \vert  \le \, n \, C^\prime \, L^{n-1} \,  
\underset{\mu_1\cdots\mu_k, z \in \partial \Gamma_n}{\rm max}\,
 {|f_{\mu_1\cdots\mu_k}(z+x_0)|} 
\end{equation}
for some constant $C^\prime$ independent of $f(x)$ and $L$. 

In the cohomological analysis of chiral anomaly,  
we encounter tensor fields which are norm-bounded 
with a reference point $x_0 \in \Gamma_n$,
an integer $p$ and a localization range $\varrho$
as 
\begin{equation}
\| f \|_{x_0,p,\varrho}  \le  C_1
\end{equation}
for a constant $C_1$ independent of $L$ (and also the gauge field).
This locality property holds true for the tensor fields which are 
obtained from the chiral anomaly $q(x)$ defined with  the overlap Dirac 
operator\cite{Neuberger:1997fp,Neuberger:1998wv,Hernandez:1998et}
by the differentiation with respect to the link variables. 
For such tensor fields, the lemma \ref{sec:poincare-lemma-on-finite-lattice}a, b
imply that 
the form $g(x)$ should satisfy the bound
\begin{equation}
\| g \|_{x_0,p,\varrho}  \le  C_2 , 
\end{equation}
and the form $\Delta f(x)$ should satisfy the more stringent bound
than eq.~(\ref{eq:bound-delta-f})
\begin{equation}
\vert \Delta f_{\mu_1\cdots\mu_{k}}(x) \vert
\le C_3  \,   L^{p} \,  {\rm e}^{-L/2\varrho}  
\end{equation}
for certain constants $C_2, C_3$ which do not depend on $L$
(and also the gauge field).
Therefore $g(x)$ has the same locality property as that of $f(x)$, 
and
$\Delta f(x)$ is a small finite-volume correction of order O(${\rm e}^{-L/2\varrho})$. 

\section{Vector potentials on the finite lattice}
\label{sec:vector-potential-on-finiteV}

In this section, we examine the parametrization of
the gauge fields on the finite lattice 
in terms of vector potentials.
We restrict ourselves to the case of four dimensions. 
It is straightforward to extend the following discussion 
to other dimensions.

In the course of the argument of the local cohomology problem,
it plays a crucial role to introduce the vector potential
which has the one-to-one correspondence to the original link 
variable. In this respect, 
an important point is that
the locality properties of gauge invariant fields should be
the same independently of whether they are considered
to be functions of the link variable or the vector 
potential. 
Similar to the lemma 5.1 of \cite{Luscher:1998kn},
the following lemma shows that for 
$\tilde U(x,\mu)$, the actual local 
and dynamical degrees of freedom in a given topological sector, 
it is possible to establish the one-to-one correspondence
to a {\it periodic} vector potential with the desired locality
properties on the finite lattice. 

\vspace{1em}
\noindent{\bf Lemma \ref{sec:vector-potential-on-finiteV}} \ \ There
exists a periodic vector potential
$\tilde A_\mu(x)$  such that 
\begin{eqnarray}
\tilde U(x,\mu) &=& {\rm e}^{i \tilde A_\mu(x)}, \\
F_{\mu\nu}(x) \,&=& \partial_\mu \tilde A_\nu(x) - \partial_\nu 
\tilde A_\mu(x) + \frac{2\pi m_{\mu\nu}}{L^2}, 
\end{eqnarray}
\begin{equation}
\left\{ 
\begin{array}{ll}
\vert \tilde A_\mu(x) \vert \le \pi(1+4 \| x \|) &\ \ 
{\rm for} \ \ x_\nu \not =  \frac{L}{2}-1  \ (\nu=1,2,3),\\
\vert \tilde A_\mu(x) \vert \le \pi(1 + 6L^2) & \ \ {\rm otherwise}.
\end{array}
\right.
\end{equation}
\noindent Moreover, if $\tilde A_\mu^\prime(x)$ is any other
field with these properties we have
\begin{equation}
\tilde A_\mu^\prime(x) = \tilde A_\mu(x)+ \partial_\mu 
\omega(x), 
\end{equation}
where the gauge function $\omega(x)$ takes values that are 
integer multiples of $2\pi$. 

\vspace{1em}
\noindent {\sl Proof:} \ \  We introduce a vector 
potential 
\begin{equation}
\tilde a_\mu(x) = \frac{1}{i} \ln \tilde U(x,\mu), \qquad
-\pi < \tilde a_\mu(x) \le \pi
\end{equation}
and then note that
\begin{equation}
F_{\mu\nu}(x)=\partial_\mu \tilde a_\nu(x) - \partial_\nu 
\tilde a_\mu(x) 
+ \frac{2\pi m_{\mu\nu}}{L^2}
+ 2\pi \tilde n_{\mu\nu}(x),
\end{equation}
where $\tilde n_{\mu\nu}(x)$ is an anti-symmetric tensor
field with integer values which satisfies 
\begin{eqnarray}
&& \partial_{[\rho} \tilde n_{\mu\nu]}(x)= 0 , \\
&& \sum_{s,t=0}^{L-1} \tilde n_{\mu\nu}(x+s\hat\mu+t\hat\nu)=0.
\end{eqnarray}
The Bianchi identity of $\tilde n_{\mu\nu}(x)$ follows from 
the Bianchi identity of $F_{\mu\nu}(x)$ which holds true for 
$\epsilon < \pi/3$.

We now construct a {\it periodic} integer vector field 
$\tilde m_\mu(x)$ such that 
$\partial_\mu \tilde m_\nu-\partial_\nu \tilde m_\mu = \tilde
n_{\mu\nu}$. For this purpose, we try to impose a complete axial 
gauge where
$\tilde m_1(x)=0$, 
$\tilde m_2(x)\vert_{x_1=0}=0$, 
$\tilde m_3(x)\vert_{x_1=x_2=0}=0$, 
$\tilde m_4(x)\vert_{x_1=x_2=x_3=0}=0$ 
and to obtain the non-zero components of the field by solving
\begin{equation}
\partial_\mu \tilde m_\nu(x) = \tilde n_{\mu\nu}(x) \ \  at \ \ 
x_1=\cdots = x_{\mu-1} = 0 
\end{equation}
for $\mu=3,2,1$ (in this order) and $\nu > \mu$. However, 
the resulted  
vector potential is not periodic.
Let us denote the restriction of the solution on to $\Gamma_4$
by $m_\mu(x)$,
\begin{equation}
m_\mu(x)= 
- \sum_{\nu < \mu} \sum_{t_\nu=0}^{x_\nu-1}{}^\prime \, 
\left.  \tilde n_{\mu\nu}(z^{(\nu)}) \right \vert_{x_1=\cdots=x_{\nu-1}=0}
\end{equation}
where $x \in \Gamma_4$, 
$z^{(\nu)}=(x_1,\cdots,t_\nu,\cdots)$
and 
\begin{equation}
\sum_{t_i =0}^{x_i-1}{}^\prime f(x)
=\left\{ 
\begin{array}{ll}
\sum_{t_i=0}^{x_i-1} f(x) & (x_i \ge 1 )\\
0 & (x_i =0 )\\
\sum_{t_i=x_i}^{-1} (-1) f(x) & (x_i \le -1 )
\end{array}
 \right. .
\end{equation}
Although it satisfies the bound $\vert m_\mu(x) \vert \le
2 \| x \|$, it only satisfies
\begin{equation}
\tilde n_{\mu\nu}= \partial_\mu m_\nu-\partial_\nu m_\mu 
+ \Delta \tilde n_{\mu\nu},
\end{equation}
\begin{equation}
\Delta \tilde n_{\mu\nu}(x)= 
\delta_{x_\mu,L/2-1}
\sum_{\tilde t_\mu=0}^{L-1}
\left.  \tilde n_{\mu\nu}(z^{(\mu,\nu)}) \right \vert_{x_{\nu+1}=\cdots=0},
\end{equation}
where $\nu > \mu$ and 
$\tilde t_\mu = t_\mu$ mod $L$. 
We note that $\Delta \tilde n_{\mu\nu}(x)$ has the support on the 
boundary of $\Gamma_4$. We then use the lattice counterpart
of the lemma 9.2 in \cite{Luscher:1998du}, to obtain the
periodic  integer vector potential $\Delta m_\mu(x)$ which solve
$\partial_\mu \Delta m_\nu-\partial_\nu \Delta m_\mu = 
\Delta \tilde n_{\mu\nu}$, 
\begin{eqnarray}
\Delta m_\mu(x) =-
\delta_{x_\mu,L/2-1}
\sum_{\nu > \mu}
\sum_{\tilde t_\mu=0}^{L-1}
\sum_{t_\nu=0}^{x_\nu-1}{}^\prime \, 
\left.  \tilde n_{\mu\nu}(z^{(\mu,\nu)}) \right \vert_{x_{\nu+1}=\cdots=0}.
\end{eqnarray}
The desired periodic integer vector potential 
$\tilde m_\mu(x)$ is now obtained by 
$\tilde m_\mu(x) = m_\mu(x) + \Delta m_\mu(x)$, 
which satisfies the bound
\begin{equation}
\left\{ 
\begin{array}{ll}
\vert \tilde m_\mu(x) \vert \le 2 \| x \| &\ \ 
{\rm for} \ \ x_\nu \not =  L/2-1  \ (\nu=1,2,3),\\
\vert \tilde m_\mu(x) \vert \le 3L^2 & \ \ {\rm otherwise}.
\end{array}
\right.
\end{equation}
and the
vector potential which represents the link variable 
$\tilde U(x,\mu)$ is obtained by
\begin{equation}
\label{eq:vector-potential-finite-lattice}
\tilde A_\mu(x) = \tilde a_\mu(x) + 2 \pi \tilde m_\mu(x). 
\end{equation}
Thus lemma establishes a one-to-one correspondence between the 
admissible 
fields $\tilde U(x,\mu)\times V_{[m]}(x,\mu)$
and the 
vector fields $\tilde A_\mu(x)$
with field tensor 
$\partial_\mu \tilde A_\nu - \partial_\nu \tilde A_\mu
=F_{\mu\nu}(x)-2\pi m_{\mu\nu}/L^2$ where 
$F_{\mu\nu}(x)$ is bounded by $\epsilon$.

Locality property of this mapping can be argued
as in the case of the lemma 5.1 in \cite{Luscher:1998kn}.
A gauge invariant field composed from the link 
variables $\tilde U(x,\mu)\, V_{[m]}(x,\mu)$ may 
be regarded as a gauge invariant field
depending the vector potential $\tilde A_\mu(x)$ % and vice versa.
and also on 
the magnetic fluxes $m_{\mu\nu}$.
The locality properties of the gauge invariant fields are 
the same independently of whether they are considered
to be functions of the link variables or the vector 
potential. Since the mapping 
\begin{equation}
\tilde A_\mu(x) \rightarrow \tilde U(x,\mu)
={\rm e}^{i \tilde A_\mu(x) } 
\end{equation}
is manifestly local, this is immediately clear
if one starts with a field composed from the link
variables. In the other direction, starting
from a gauge invariant local field $\phi(y)$ 
depending on the vector potential (and also on 
the magnetic fluxes $m_{\mu\nu}$), the key 
observation is that one is free to 
change the gauge of the integer field $\tilde m_\mu(x)$
in eq.~(\ref{eq:vector-potential-finite-lattice}).
In particular, we may impose a complete axial gauge
taking the point $y$ as the origin. Around $y$ the 
vector potential is then locally constructed from the given 
link field and $\phi(y)$ thus maps to a local function
of the link variables residing there.\footnote{
As shown in \cite{Luscher:1998du},  $\tilde U(x,\mu)$
can be parametrized uniquely by 
the Wilson lines $w_\mu$, the transverse vector potential
$A_\mu^T(x)$  and the gauge function $\Lambda(x)$. 
This parametrization of 
the link variable, however,  does not possess the 
desired locality properties and  does not suit for
our purpose. 
}

\section{Cohomological analysis of topological fields on the finite lattice}
\label{sec:chiral-anomaly-in-finite-lattice}

In this section, 
equipped with the Poincar\'e lemma reformulated for the finite lattice and
the periodic and bounded vector potential representation of the link variables,
we now perform a cohomological analysis
of topological fields directly on the finite lattice. 
We consider a gauge-invariant local field  $q(x)$ on the finite volume lattice
which is topological in the sense that 
\begin{equation}
  \sum_{x \in \Gamma_4} q(x) = \text{integer}
\end{equation}
and also that 
\begin{equation}
\sum_{x \in \Gamma_4}  \delta q(x) = 0 
\end{equation}
for any local variation of the gauge field. 
$q(x)$ can be  separated into two parts,  
the part defined in the infinite lattice $q_\infty(x)$
and the part of finite-volume correction $\Delta_L q(x)$ as
\begin{equation}
%\label{eq:q-separation}
q(x)=q_\infty(x) + \Delta q_\infty(x) . 
\end{equation}
$q_\infty(x)$ is a gauge-invariant local field defined in the 
infinite lattice, but restricted with periodic gauge fields.
We assume that this part is also  topological 
in the sense that 
\begin{equation}
\sum_{x \in \mathbb{Z}^4}  \delta q_\infty(x) = 0 , 
\end{equation}
for any local variation of the gauge field (not restricted to periodic gauge fields). 
On the other hand,  $\Delta q_\infty(x)$ satisfies the bound
\begin{equation}
\vert \Delta q_\infty(x) \vert \le  \kappa L^\sigma {\rm e}^{-L/2\varrho} .
\end{equation}
for a constant $\kappa$ and an integer $\sigma$.  $\varrho$ is the
localization range of $q(x)$. 
This property of $q(x)$ allows us to 
relate our result obtained directly on the finite lattice
to that obtained through the cohomological analysis in the infinite 
lattice\cite{Luscher:1998kn,Igarashi:2002zz}. 
All the above properties of the topological field $q(x)$
are satisfied for the chiral anomaly given in  terms of 
the overlap Dirac operator.

\subsection{Cohomological analysis of topological field on the finite lattice}

Let $q(x)$ be a gauge-invariant local field 
on the finite lattice which is topological in the sense that 
\begin{equation}
\label{eq:chiral-anomaly-topological}
\sum_{x \in \Gamma_4}  \delta q(x) = 0 , 
\end{equation}
for any local variation of the gauge field. 
Our aim is then to establish 
\begin{eqnarray}
\label{eq:chiral-anomaly-result}
q(x)= q_{ [m]}(x) %\frac{Q}{L^4} %
+ \phi_{[m]\mu\nu}(x) \tilde F_{\mu\nu}(x) 
+
\gamma_{[m,w]} \epsilon_{\mu\nu\rho\sigma}
\tilde F_{\mu\nu}(x) \, \tilde F_{\rho\sigma}(x+\hat \mu+\hat \nu)
+ \partial_\mu^\ast  \tilde k_\mu(x), \ \ %\nonumber\\
\end{eqnarray}
where $q_{ [m]}(x)$ denotes the same field 
for the configuration $V_{[m]}(x,\mu)$, 
$\phi_{[m]\mu\nu}(x)$
is a gauge invariant functional of $V_{[m]}(x,\mu)$,  
$\gamma_{[m,w]}$ is a constant which may depend on $V_{[m]}(x,\mu)$ and
a constant Wilson line,
and $\tilde k_\mu(x)$ is a gauge invariant local current. 
The first step of the proof of 
eq.~(\ref{eq:chiral-anomaly-result}) is the following lemma, 
which corresponds to the lemma 6.1 in \cite{Luscher:1998kn}.

\vspace{1em}
\noindent{\bf Lemma \ref{sec:chiral-anomaly-in-finite-lattice}.1} \ \ 
There exist gauge invariant local fields $\phi_{\mu\nu}(x)$ and
$h_\mu(x)$ such that 
\begin{eqnarray}
&& \phi_{\mu\nu}(x) = - \phi_{\nu\mu}(x), \qquad
\partial_\mu^\ast  \phi_{\mu\nu}(x)=0, \\
&& q(x)=q_{ [m]}(x) + \phi_{\mu\nu}(x)\, \tilde
F_{\mu\nu}(x)  +\partial_\mu^\ast h_\mu(x).
\end{eqnarray}

\vspace{1em}
\noindent{Proof:} \ \  The vector potential $\tilde A_\mu(x)$
represents an admissible field through 
${\rm e}^{i \tilde A_\mu(x)} \times V_{[m]}(x,\mu)$ and 
the associated field tensor $F_{\mu\nu}(x) 
= \partial_\mu \tilde A_\nu(x)- \partial_\nu \tilde A_\mu(x)
+ \frac{2\pi m_{\mu\nu}}{L^2}$
is hence bounded by $\epsilon$. It is straightforward to 
check that this property is preserved if the potential is scaled
by a factor $t$ in the range $0 \le t \le 1$, i.e. we can  contract
the vector potential to zero without leaving the space of 
admissible fields. Differentiation and integration with
respect to $t$ then yields
\begin{equation}
\label{eq:chiral-anomaly-in-current}
q(x) = q_{ [m]}(x) + \sum_{y \in \Gamma_4} 
j_\nu(x,y) \tilde A_\nu(y), 
\end{equation}
where 
\begin{equation}
j_\nu(x,y) = \int_0^1 dt \left(
\frac{\partial q(x)}{\partial 
\tilde A_\nu(y)} \right)_{\tilde A \rightarrow t \tilde A} .
\end{equation}
As a function of the gauge fields, 
the current $j_\nu(x,y)$ has the same
locality properties as the topological field. 

Since the gauge group is abelian, the derivative 
of a gauge invariant field with respect to the vector 
potential is gauge invariant and the same is hence true
for $j_\nu(x,y)$. Performing an infinitesimal gauge 
transformation in eq.~(\ref{eq:chiral-anomaly-in-current}),
it then follows that 
\begin{equation}
j_\nu(x,y) \overleftarrow{\partial_\nu^\ast} = 0.
\end{equation}
Here and below the convention is adopted that a difference operator
refers to $x$ or $y$ depending on whether it appears on the left
or the right of an expression. 

The lemma \ref{sec:poincare-lemma-on-finite-lattice}a now 
allows us to conclude that there exists
a gauge invariant anti-symmetric tensor field 
$\theta_{\mu\nu}(x,y)$ such that
\begin{eqnarray}
\label{eq:j-to-theta}
j_\nu(x,y) =\theta_{\nu\mu}(x,y) \overleftarrow{\partial_\mu^\ast} 
+ \Delta j_\nu(x,y),\qquad
\vert \Delta j_\nu(x,y) \vert \le \kappa_1 
L^{\sigma_1} {\rm e}^{-L/2\rho}.\label{eq:current-to-anti-symm-tensor}
\end{eqnarray}
As explained in Sec.~\ref{sec:poincare-lemma-on-finite-lattice}, 
the construction of this field involves a reference point 
$x_0$ which is here taken to be $x$. This choice ensures that
$\theta_{\mu\nu}(x,y)$ has the same locality properties as $j_\nu(x,y)$.
$\Delta j_\nu(x,y)$ is a small field which satisfies the exponential bound. 
In the following, 
the symbol $\Delta$ should be understood to 
denote such exponentially small fields satisfying certain similar bounds. 

When eq.~(\ref{eq:current-to-anti-symm-tensor}) is 
inserted in eq.~(\ref{eq:chiral-anomaly-in-current}), 
a partial summation yields
\begin{eqnarray}
q(x)=q_{[m]}(x) +\frac{1}{2} \sum_{y\in \Gamma_4}
\theta_{\mu\nu}(x,y)\, \tilde F_{\mu\nu}(y)
+ \sum_{y \in \Gamma_4} 
\Delta j_\nu(x,y) \tilde A_\nu(y).\label{eq:chiral-anomaly-in-theta}
\end{eqnarray}
This may be rewritten in the form
\begin{eqnarray}
q(x)=q_{[m]}(x) 
+ \tilde \phi_{\mu\nu}(x) \, \tilde F_{\mu\nu}(x)
+ \frac{1}{2} \sum_{y\in \Gamma_4}
\eta_{\mu\nu}(x,y)\, \tilde F_{\mu\nu}(y) 
\label{eq:chiral-anomaly-in-phi-prime}
+ \sum_{y \in \Gamma_4} 
\Delta j_\nu(x,y) \tilde A_\nu(y),\ \ 
\end{eqnarray}
where the new fields are given by
\begin{eqnarray}
&& \tilde \phi_{\mu\nu}(x) = \frac{1}{2}\sum_{z \in \Gamma_4}
\theta_{\mu\nu}(z,x) \qquad
\label{eq:tilde-phi}\\
&& \eta_{\mu\nu}(x,y) = \theta_{\mu\nu}(x,y) - 
   \delta_{x,y} \sum_{z \in \Gamma_4} \theta_{\mu\nu}(z,y).\qquad
\end{eqnarray}
Both of them are gauge invariant and anti-symmetric in the 
indices $\mu,\nu$. Moreover, taking the locality properties
of $\theta_{\mu\nu}(x,y)$ into account, it is easy
to prove that $\tilde \phi_{\mu\nu}(x)$ is a local 
field. 

Using eq.~(\ref{eq:current-to-anti-symm-tensor})
and 
the topological properties of $q(x)$
one obtains
\begin{equation}
\label{eq:premise-to-redefine-phi}
\partial_\mu^\ast \tilde \phi_{\mu\nu}(x)
= \frac{1}{2}\sum_{z \in \Gamma_4}  \Delta j_\nu(z,x).
\end{equation}
Since the r.h.s of eq.~(\ref{eq:premise-to-redefine-phi}) is an
exact form, 
it satisfies the premise of the lemma \ref{sec:poincare-lemma-on-finite-lattice}b. Therefore we may apply the 
lemma \ref{sec:poincare-lemma-on-finite-lattice}b to obtain
\begin{equation}
\label{eq:Delta-Phi}
\frac{1}{2}\sum_{z \in \Gamma_4}  \Delta j_\nu(z,x)
= \partial_\mu^\ast\Delta \Phi_{\mu\nu}(x). 
\end{equation}
We may then define another tensor field
\begin{equation}
\label{eq:phi}
\phi_{\mu\nu}(x) \equiv \tilde \phi_{\mu\nu}(x)
-\Delta \Phi_{\mu\nu}(x), 
\end{equation}
which satisfies
\begin{equation}
\partial_\mu^\ast \phi_{\mu\nu}(x) = 0.
\end{equation}
From the redefinition of $\tilde \phi_{\mu\nu}(x)$, 
eq.~(\ref{eq:chiral-anomaly-in-phi-prime}) may 
be rewritten in the form
\begin{eqnarray}
q(x)=q_{[m]}(x) 
+ \phi_{\mu\nu}(x) \, \tilde F_{\mu\nu}(x)
+ \frac{1}{2} \sum_{y\in \Gamma_4}
\eta_{\mu\nu}(x,y)\, \tilde F_{\mu\nu}(y) 
+ \Delta q(x), \label{eq:chiral-anomaly-in-phi-redefinition}
\end{eqnarray}
where
\begin{eqnarray}
\Delta q(x) =\Delta \Phi_{\mu\nu}(x) \tilde F_{\mu\nu}(x)
+\sum_{y \in \Gamma_4} 
\Delta j_\nu(x,y) \tilde A_\nu(y).
\end{eqnarray}

As for the fields $\eta_{\mu\nu}(x,y),\Delta q(x)$ we note that
\begin{equation}
\sum_{x \in \Gamma_4} \eta_{\mu\nu}(x,y) = 0, \qquad 
\sum_{x \in \Gamma_4} \Delta q(x) = 0
\end{equation}
and the lemma \ref{sec:poincare-lemma-on-finite-lattice}a 
(or \ref{sec:poincare-lemma-on-finite-lattice}b) may hence be applied again. 
This leads to the representation
\begin{equation}
\eta_{\mu\nu}(x,y) = \partial_\lambda^\ast
\tau_{\lambda\mu\nu}(x,y), 
\qquad \Delta q(x) = \partial_\lambda^\ast \Delta h_{\lambda}(x)
\end{equation}
in terms of new fields $\tau_{\lambda\mu\nu}(x,y),\Delta h_{\lambda}(x)$.
Then we define
\begin{equation}
h_\mu(x) = \frac{1}{2}\sum_{y \in \Gamma_4} 
\tau_{\mu\nu\rho}(x,y) \tilde F_{\nu\rho}(y) + \Delta h_{\mu}(x).
\end{equation}

With these results, eq.~(\ref{eq:chiral-anomaly-in-phi-redefinition}) may 
be finally rewritten in the form
\begin{eqnarray}
\label{eq:chiral-anomaly-in-first-step}
q(x)=q_{[m]}(x) 
+ \phi_{\mu\nu}(x) \, \tilde F_{\mu\nu}(x)
+ \partial_\mu^\ast h_\mu(x)
\end{eqnarray}
and the lemma has thus been proved.

In the second step of the proof of 
eq.~(\ref{eq:chiral-anomaly-result}) we determine the general form of 
the field $\phi_{\mu\nu}(x)$ using no other properties
than those stated in 
lemma \ref{sec:chiral-anomaly-in-finite-lattice}.1. 
This step corresponds to the lemma 6.2 in \cite{Luscher:1998kn}.

\vspace{1em}
\noindent{\bf Lemma \ref{sec:chiral-anomaly-in-finite-lattice}.2} \ \
There exists a gauge invariant, local and totally anti-symmetric
tensor field $t_{\lambda\mu\nu}(x)$ such that 
\begin{eqnarray}
\phi_{\mu\nu}(x) = \phi_{[m] \mu\nu}(x)
+ \gamma_{[m,w]} \epsilon_{\mu\nu\rho\sigma} 
\tilde F_{\rho\sigma}(x+\hat\mu+\hat\nu)
+ \partial_\lambda^\ast t_{\lambda\mu\nu}(x)
+\Delta \phi_{\mu\nu}(x),  \qquad
\end{eqnarray}
where $\phi_{[m] \mu\nu}(x)$ is the value of $\phi_{\mu\nu}(x)$ at 
$V_{[m]}(x,\mu)$, and $\gamma_{[m,w]}$ is a constant which may depends
on $V_{[m]}(x,\mu)$ and a constant Wilson line. $\Delta
\phi_{\mu\nu}(x)$ satisfies the bound
$\vert \Delta \phi_{\mu\nu}(x) \vert 
\le \kappa_2  L^{\sigma_2} {\rm e}^{-L/2\rho}$.

\vspace{1em}
\noindent{Proof:} \ \  Proceeding as in the proof of lemma
\ref{sec:chiral-anomaly-in-finite-lattice}.1, it is straightforward
to derive a representation analogous to 
eq.~(\ref{eq:chiral-anomaly-in-theta})
for  the field $\phi_{\mu\nu}(x)$. Only the locality and
gauge invariance of the field are required for this
and one ends up with the expression
\begin{eqnarray}
\phi_{\mu\nu}(x) = \phi_{[m] \mu\nu}(x)
+\frac{1}{2} \sum_{y \in \Gamma_4} 
\tilde \xi_{\mu\nu\rho\sigma}(x,y)
\tilde F_{\rho\sigma}(y)
+ \sum_{y \in \Gamma_4} \Delta j_{\mu\nu\rho}(x,y) \, 
\tilde A_\rho(y), \quad\label{eq:chiral-anomaly-phi-in-xi}
\end{eqnarray}
where $\phi_{[m] \mu\nu}(x)$ is the value of 
$\phi_{\mu\nu}(x)$ at $V_{[m]}(x,\mu)$
and the new fields $\tilde\xi_{\mu\nu\rho\sigma}(x,y)$ and
$\Delta j_{\mu\nu\rho}(x,y)$ are defined through
\begin{eqnarray}
j_{\mu\nu\rho}(x,y) = \int_0^1 dt \left(
\frac{\partial \phi_{\mu\nu}(x)}{\partial 
\tilde A_\rho(y)} \right)_{\tilde A \rightarrow t \tilde A},\quad j_{\mu\nu\rho}(x,y) 
\overleftarrow{\partial_\rho^\ast} =0, 
\end{eqnarray}
\begin{equation}
\label{eq:phi-current-xi}
j_{\mu\nu\rho}(x,y)
=\tilde 
\xi_{\mu\nu\rho\sigma}(x,y)
\overleftarrow{\partial_\sigma^\ast} +\Delta j_{\mu\nu\rho}(x,y). \hspace{2.8cm} 
\hspace{1.2mm}
\end{equation}
As a function of the gauge fields, 
these new fields appearing 
in eq.~(\ref{eq:chiral-anomaly-phi-in-xi}) 
are gauge invariant and have the same locality properties
as the current $j_{\mu\nu\rho}(x,y)$, that is, 
as the field $\phi_{\mu\nu}(x)$. 
From eq.~(\ref{eq:phi-current-xi}) it follows that 
$\tilde \xi_{\mu\nu\rho\sigma}(x,y)$ satisfies
\begin{eqnarray}
&&\tilde \xi_{\mu\nu\rho\sigma} 
= - \tilde \xi_{\nu\mu\rho\sigma} =
-\tilde \xi_{\mu\nu\sigma\rho}, \qquad\qquad\\ 
&&\partial_\mu^\ast \tilde \xi_{\mu\nu\rho\sigma}(x,y)
\label{eq:chiral-anomaly-xi-properties}
\overleftarrow{\partial_\sigma^\ast} 
=-\partial_\mu^\ast 
\Delta j_{\mu\nu\rho}(x,y). \qquad\qquad
\end{eqnarray}
Since the r.h.s of the second equation of 
eq.~(\ref{eq:chiral-anomaly-xi-properties}) is 
an exact form in terms of $y$, it satisfies the premise of the 
lemma \ref{sec:poincare-lemma-on-finite-lattice}b. 
Therefore we may apply the lemma \ref{sec:poincare-lemma-on-finite-lattice}b to 
obtain
\begin{equation}
\label{eq:lemma-delta-j3}
-\partial_\mu^\ast 
\Delta j_{\mu\nu\rho}(x,y) 
= \partial_\mu^\ast 
\Delta \Xi_{\mu\nu\rho\sigma}(x,y) 
\overleftarrow{\partial_\sigma^\ast}.\qquad
\end{equation}
Here we note that 
$\partial_\mu^\ast$ can be 
extracted explicitly in the r.h.s. of eq.(\ref{eq:lemma-delta-j3}).
This is because 
$\partial_\mu^\ast \Delta j_{\mu\nu\lambda}(x,y)$ is already an
exponentially small field 
and when applying the lemma \ref{sec:poincare-lemma-on-finite-lattice}b with 
respect to $y$, the reference
point $x_0$ is not necessarily identified with $x$.
We may then define another tensor field 
\begin{equation}
\xi_{\mu\nu\rho\sigma}(x,y)
=\tilde \xi_{\mu\nu\rho\sigma}(x,y)
-\Delta \Xi_{\mu\nu\rho\sigma}(x,y),
\end{equation}
which satisfies
\begin{equation}
\label{eq:xi-tilde-property}
\partial_\mu^\ast \xi_{\mu\nu\rho\sigma}(x,y)
\overleftarrow{\partial_\sigma^\ast}  =0.
\end{equation}

Applying the lemma \ref{sec:poincare-lemma-on-finite-lattice}a to 
eq.~(\ref{eq:xi-tilde-property})
we have
\begin{equation}
\label{eq:partial-xi-tilde-lemma}
\partial_\mu^\ast \xi_{\mu\nu\rho\sigma}(x,y)
= \tilde \upsilon_{\nu\rho\sigma\tau}(x,y)
\overleftarrow{\partial_\tau^\ast} 
+ 
\Delta\{ \partial_\mu^\ast \xi_{\mu\nu\rho\sigma}\}(x,y),
\end{equation}
\begin{equation}
\label{eq:chiral-anomaly-v-properties}
\partial_\nu^\ast
\tilde \upsilon_{\nu\rho\sigma\tau}(x,y)
\overleftarrow{\partial_\tau^\ast}  =
- \partial_\nu^\ast
\Delta\{ \partial_\mu^\ast
\xi_{\mu\nu\rho\sigma}\}(x,y) .\hspace{1.5cm} 
\end{equation}
From eq.~(\ref{eq:chiral-anomaly-v-properties}),
we may again apply the lemma \ref{sec:poincare-lemma-on-finite-lattice}b to obtain
\begin{equation}
-\partial_\nu^\ast
\Delta\{ \partial_\mu^\ast
\xi_{\mu\nu\rho\sigma}\}(x,y)
=\partial_\nu^\ast
\Delta \Upsilon_{\nu\rho\sigma\tau}(x,y)
\overleftarrow{\partial_\tau^\ast} .
\end{equation}
We may then define another tensor field
\begin{equation}
\upsilon_{\nu\rho\sigma\tau}(x,y)
=\tilde \upsilon_{\nu\rho\sigma\tau}(x,y)
-\Delta\Upsilon_{\nu\rho\sigma\tau}(x,y), 
\end{equation}
which satisfies
\begin{equation}
\label{eq:upsilon-tilde-property}
\partial_\nu^\ast
\upsilon_{\nu\rho\sigma\tau}(x,y)
\overleftarrow{\partial_\tau^\ast} 
=0.
\end{equation}

Applying the lemma \ref{sec:poincare-lemma-on-finite-lattice}a to 
eq.~(\ref{eq:upsilon-tilde-property})
we have
\begin{equation}
\label{eq:partial-upsilon-tilde-lemma}
\partial_\nu^\ast
\upsilon_{\nu\rho\sigma\tau}(x,y) 
= \tilde \omega_{\rho\sigma\tau\lambda}(x,y)
\overleftarrow{\partial_\lambda^\ast} 
+ \Delta\{\partial_\nu^\ast
\upsilon_{\nu\rho\sigma\tau}\}(x,y), 
\end{equation}
\begin{equation}
\label{eq:omega-property}
\sum_{x \in \Gamma_4} \tilde \omega_{\rho\sigma\tau\lambda}(x,y) 
\overleftarrow{\partial_\lambda^\ast}
=-\sum_{x \in \Gamma_4} \Delta\{\partial_\nu^\ast
\upsilon_{\nu\rho\sigma\tau}\}(x,y).\ \ \ \ \ \ 
\end{equation}
Since the r.h.s of eq.~(\ref{eq:omega-property}) satisfies
the premise of lemma \ref{sec:poincare-lemma-on-finite-lattice}b.
Repeatedly, we may apply the lemma \ref{sec:poincare-lemma-on-finite-lattice}b 
to obtain
\begin{equation}
-\sum_{x \in \Gamma_4} \Delta\{\partial_\nu^\ast
\upsilon_{\nu\rho\sigma\tau}\}(x,y)
=
\Delta \Omega_{\rho\sigma\tau\lambda}(y)
\overleftarrow{\partial_\lambda^\ast}.
\end{equation}
We may then define another tensor field
\begin{equation}
\omega_{\rho\sigma\tau\lambda}(x,y)
=\tilde \omega_{\rho\sigma\tau\lambda}(x,y)
-\delta_{x,y}\, \Delta \Omega_{\rho\sigma\tau\lambda}(y), 
\end{equation}
which satisfies
\begin{equation}
\sum_{x \in \Gamma_4} \omega_{\rho\sigma\tau\lambda}(x,y)
\overleftarrow{\partial_\lambda^\ast} = 0.
\end{equation}
An immediate consequence of the last equation is that
\begin{equation}
2 \gamma_{[m,w]} \, \epsilon_{\rho\sigma\tau\lambda}
=\sum_{z \in \Gamma_4} \omega_{\rho\sigma\tau\lambda}(z,x)
\end{equation}
is independent of $x$. In view of the locality properties 
of the expression, a dependence on the vector potential is 
almost excluded except the Wilson line $w_\mu$.
It also depends on 
the magnetic flux $m_{\mu\nu}$ and the size of the lattice $L$.
We will discuss this point later
in relation to the anomaly cancellation.

Another application of the lemma \ref{sec:poincare-lemma-on-finite-lattice}a 
(or \ref{sec:poincare-lemma-on-finite-lattice}b) now implies that 
\begin{equation}
\omega_{\rho\sigma\tau\lambda}(x,y)
= 2\gamma_{[m,w]} \, \epsilon_{\rho\sigma\tau\lambda} \, \delta_{x,y} 
+\partial_\nu^\ast \varphi_{\nu\rho\sigma\tau\lambda}(x,y)
\end{equation}
for some vector field (with
respect to $x$), $\varphi_{\nu\rho\sigma\tau\lambda}(x,y)$.  
If we define
\begin{equation}
\check \upsilon_{\nu\rho\sigma\tau}(x,y) =
\upsilon_{\nu\rho\sigma\tau}(x,y) -
\varphi_{\nu\rho\sigma\tau\lambda}(x,y)
\overleftarrow{\partial_\lambda^\ast}, 
\end{equation}
it is then straightforward to prove the relations
\begin{eqnarray} 
\label{eq:xi-properties-with-hat}
\partial_\mu^\ast \xi_{\mu\nu\rho\sigma}(x,y)
=\check \upsilon_{\nu\rho\sigma\tau}(x,y)
\overleftarrow{\partial_\tau^\ast} 
+ 
\Delta\{ \partial_\mu^\ast \xi_{\mu\nu\rho\sigma}\}(x,y), \quad
\end{eqnarray}
\begin{equation} 
\label{eq:v-properties-with-hat}
\partial_\nu^\ast
\check \upsilon_{\nu\rho\sigma\tau}(x,y) 
=\left\{
2 \gamma_{[m,w]} \epsilon_{\rho\sigma\tau\lambda} \, \delta_{x,y}
\right\}\overleftarrow{\partial_\lambda^\ast}
+\left\{\delta_{x,y}\Delta \Omega_{\rho\sigma\tau\lambda} (y) 
 \right\}\overleftarrow{\partial_\lambda^\ast} 
+ \Delta\{\partial_\nu^\ast
\upsilon_{\nu\rho\sigma\tau}\}(x,y).\quad
\end{equation}
Compared to 
eqs.~(\ref{eq:partial-xi-tilde-lemma}),
(\ref{eq:partial-upsilon-tilde-lemma})
the  important difference is that the form of the first term 
in the right hand
side of the second equation is now known precisely. 

In the next step we propagate this information to the first
equation by noting
\begin{equation}
\label{eq:delta-partial}
\delta_{x,y}\partial_\lambda^\ast = - \partial_\nu^\ast
\left\{ \delta_{\nu\lambda} \delta_{x,y-\hat\nu} \right\},\qquad
\end{equation}
\begin{equation}
\sum_{ x \in \Gamma_4}
\Delta\{\partial_\nu^\ast
\upsilon_{\nu\rho\sigma\tau}\}(x,y)
=  0 .\qquad
\end{equation}
The general solution of eq.~(\ref{eq:v-properties-with-hat}) is 
hence given by
\begin{eqnarray}
\check v_{\nu\rho\sigma\tau}(x,y) 
=  
-\delta_{\nu\lambda} \delta_{x,y-\hat\nu} 2\gamma_{[m,w]} 
\epsilon_{\rho\sigma\tau\lambda}
+\partial_\mu^\ast \theta_{\mu\nu\rho\sigma\tau}(x,y)
+ \Delta \check v_{\nu\rho\sigma\tau}(x,y),\quad\quad
\end{eqnarray}
where
$\theta_{\mu\nu\rho\sigma\tau}=-\theta_{\nu\mu\rho\sigma\tau}$.
It follows from this that the shifted field
\begin{equation}
\check \xi_{\mu\nu\rho\sigma}(x,y)=
\xi_{\mu\nu\rho\sigma}(x,y)
-\theta_{\mu\nu\rho\sigma\tau}(x,y) 
\overleftarrow{\partial_\tau^\ast}
\end{equation}
satisfies the relation
\begin{eqnarray}
\partial_\mu^\ast \check \xi_{\mu\nu\rho\sigma}(x,y)
=
-2 \gamma_{[m,w]} \delta_{\nu\lambda} \delta_{x,y-\hat\nu}
\overleftarrow{\partial_\tau^\ast} 
\epsilon_{\rho\sigma\tau\lambda}
 +\Delta\{ \partial_\mu^\ast \check \xi_{\mu\nu\rho\sigma}\}(x,y),
\quad\quad
\end{eqnarray}
where
\begin{eqnarray}
\Delta\{ \partial_\mu^\ast \check \xi_{\mu\nu\rho\sigma}\}(x,y)
=
\Delta\{ \partial_\mu^\ast \xi_{\mu\nu\rho\sigma}\}(x,y)
+\Delta \check v_{\nu\rho\sigma\tau}(x,y)
\overleftarrow{\partial_\tau^\ast}.\qquad
\end{eqnarray}
We may now again use the identity eq.~(\ref{eq:delta-partial}) and 
the lemma \ref{sec:poincare-lemma-on-finite-lattice}a 
(or \ref{sec:poincare-lemma-on-finite-lattice}b) to infer that 
\begin{eqnarray}
\check \xi_{\mu\nu\rho\sigma}(x,y)
= 2 \gamma_{[m,w]}
\epsilon_{\mu\nu\rho\sigma}\delta_{x,y-\hat\mu-\hat\nu}
+\partial_\lambda^\ast
\kappa_{\lambda\mu\nu\rho\sigma}(x,y)
+\Delta\check \xi_{\mu\nu\rho\sigma}(x,y)
\end{eqnarray}
where $\kappa_{\lambda\mu\nu\rho\sigma}(x,y)$ and 
$\Delta\check \xi_{\mu\nu\rho\sigma}(x,y)$ are another 
tensor fields.

Together with
\begin{eqnarray}
\label{eq:chiral-anomaly-phi-in-xi-check}
\phi_{\mu\nu}(x) =\phi_{[m] \mu\nu}(x)
+\frac{1}{2} \sum_{y \in \Gamma_4} 
\check \xi_{\mu\nu\rho\sigma}(x,y)
\tilde F_{\rho\sigma}(y) 
+
\frac{1}{2} \sum_{y \in \Gamma_4} 
\Delta \Xi_{\mu\nu\rho\sigma}(x,y)
\tilde F_{\rho\sigma}(y)\nonumber\\
+ \sum_{y \in \Gamma_4} \Delta j_{\mu\nu\rho}(x,y) \, 
\tilde A_\rho(y), \hspace{6cm}
\end{eqnarray}
and the definitions
\begin{eqnarray}
t_{\lambda \mu \nu}(x)\,=
\frac{1}{2}\sum_{y \in \Gamma_4} 
\kappa_{\lambda \mu\nu \rho \sigma}(x,y)
\tilde F_{\rho \sigma}(y),\hspace{5.1cm} 
\end{eqnarray}\vspace{-5mm}
\begin{eqnarray}
\Delta \phi_{\mu\nu}(x)
=\frac{1}{2} \sum_{y \in \Gamma_4} 
\Delta \check \xi_{\mu\nu\rho\sigma}(x,y)
\tilde F_{\rho\sigma}(y) 
+\frac{1}{2} \sum_{y \in \Gamma_4} 
\Delta \Xi_{\mu\nu\rho\sigma}(x,y)
\tilde F_{\rho\sigma}(y) 
\quad\nonumber\\
+ \sum_{y \in \Gamma_4} \Delta j_{\mu\nu\rho}(x,y) \, 
\tilde A_\rho(y),\hspace{5cm}
\end{eqnarray}
this proves the lemma. 

The combination of lemma \ref{sec:chiral-anomaly-in-finite-lattice}.1 and 
\ref{sec:chiral-anomaly-in-finite-lattice}.2 leads to the 
representation
\begin{eqnarray}
q(x)= q_{[m]}(x) + \phi_{[m] \mu\nu}(x)\,
\tilde F_{\mu\nu}(x)  
+\gamma_{[m,w]} \epsilon_{\mu\nu\rho\sigma} \tilde F_{\mu\nu}(x) 
\tilde F_{\rho\sigma}(x+\hat\mu+\hat\nu) +
\partial_\mu^\ast  h_\mu(x)\nonumber\\
+ \partial_\lambda^\ast t_{\lambda\mu\nu}(x) \tilde
F_{\mu\nu}(x)  
+ \Delta \phi_{\mu\nu}(x)\, \tilde F_{\mu\nu}(x).\hspace{5cm}
\end{eqnarray}
Using the anti-symmetry of the tensor field 
$t_{\lambda\mu\nu}(x)$ and the vanishing of the 
monopole current, $\epsilon_{\mu\nu\rho\sigma} \partial_\nu
\tilde F_{\rho\sigma}(x)=0$, it is easy to check that
\begin{equation}
\partial_\mu^\ast t_{\mu\nu\rho}(x) \tilde F_{\nu\rho}(x)
=\partial_\mu^\ast\left\{
t_{\mu\nu\rho}(x) \tilde F_{\nu\rho}(x+\hat\mu)\right\}.
\end{equation}
Moreover it follows that 
\begin{equation}
\sum_{x \in \Gamma_4} 
\Delta \phi_{\mu\nu}(x)\, \tilde F_{\mu\nu}(x)
=0.  
\end{equation}
Since the tensor fields in the r.h.s. are exponentially small, 
we may apply the lemma \ref{sec:poincare-lemma-on-finite-lattice}b 
with an arbitrary reference point $x_0$ to obtain
\begin{equation}
 \Delta \phi_{\mu\nu}(x)\, \tilde F_{\mu\nu}(x)
=\partial_\mu^\ast \Delta k_\mu(x) . 
\end{equation}
This proves eq.~(\ref{eq:chiral-anomaly-result})
with the definition of $\tilde k_\mu(x)$ as
\begin{equation}
\tilde k_\mu(x) =  h_\mu(x) +  t_{\mu\nu\rho}(x) \tilde F_{\nu\rho}(x+\hat\mu)
+ \Delta k_\mu(x).
\end{equation}

\subsection{Relation to the result obtained in the infinite lattice}

We will next compare our result eq.~(\ref{eq:chiral-anomaly-result})
with the result obtained through the cohomological analysis in the infinite 
lattice\cite{Luscher:1998kn,Igarashi:2002zz}. 
The later result may be summarized as follows: 
let us assume that the topological field $q(x)$ on the finite-volume lattice can be  separated into two parts,  
the part defined in the infinite lattice and the part of finite-volume correction,
\begin{equation}
\label{eq:q-separation}
q(x)=q_\infty(x) + \Delta q_\infty(x) . 
\end{equation}
$q_\infty(x)$ is the topological field
defined in the infinite lattice and with the periodic link variables. 
$\Delta q_\infty(x)$ satisfies the bound
\begin{equation}
\vert \Delta q\infty (x) \vert \le  \kappa L^\sigma {\rm e}^{-L/2\varrho} 
\end{equation}
for a constant $\kappa$ and an integer $\nu$.  $\varrho$ is the
localization range of $q(x)$. 
Then, through the cohomological analysis in the infinite lattice\cite{Luscher:1998kn}, 
the first part can be expressed as  
\begin{equation}
q_\infty(x) = 
\alpha + \beta_{\mu\nu}  F_{\mu\nu}(x) 
+\gamma \epsilon_{\mu\nu\rho\sigma}
F_{\mu\nu}(x) \, F_{\rho\sigma}(x+\hat \mu+\hat \nu)
+\partial_\mu^\ast k_{\mu \infty}(x) .  
\end{equation}
The part of the finite-volume 
correction is an exponentially small field and it cannot contribute
to the topological charge for a sufficiently large $L$. Namely we have
\begin{equation}
\sum_{x\in \Gamma_4} \Delta q_\infty (x) = 0. 
\end{equation}
Then it can be expressed 
as the total-divergence of a certain gauge-invariant current 
which is also exponentially small \cite{Igarashi:2002zz}, 
\begin{equation}
 \Delta q_\infty (x) = \partial_\mu^\ast  \Delta k_{\mu\infty}(x) . 
\end{equation}
The total topological field can be expressed as 
\begin{equation}
\label{eq:chiral-anomaly-result-infinite-lattice}
q(x) =\alpha + \beta_{\mu\nu}  F_{\mu\nu}(x) 
+ \gamma \epsilon_{\mu\nu\rho\sigma}
F_{\mu\nu}(x) \, F_{\rho\sigma}(x+\hat \mu+\hat \nu)
+\partial_\mu^\ast k_{\mu }(x) ,
\end{equation}
where 
\begin{equation}
\label{eq:current-k-infinite-lattice}
k_{\mu }(x) \equiv k_{\mu \infty}(x) + \Delta k_{\mu\infty}(x).
\end{equation}

Since 
$F_{\mu\nu}(x)=\frac{2\pi m_{\mu\nu}}{L^2}+ \tilde F_{\mu\nu}(x)$, 
the cohomologically non-trivial part of the above result may be rewritten as
\begin{eqnarray}
\label{eq:chiral-anomaly-result-infinite-lattice-tildeF}
q(x) &=& \alpha + \beta_{\mu\nu} \frac{2\pi m_{\mu\nu}}{L^2}
+ \gamma \epsilon_{\mu\nu\rho\sigma} 
\frac{2\pi m_{\mu\nu}}{L^2} \frac{2\pi m_{\rho\sigma}}{L^2}
\nonumber\\
&&\quad + \beta_{\mu\nu} \tilde F_{\mu\nu}(x) 
+2\, \gamma \epsilon_{\mu\nu\rho\sigma}
\frac{2\pi m_{\rho\sigma}}{L^2} \, 
\tilde F_{\mu\nu}(x)
\nonumber\\
&&\qquad +\gamma \epsilon_{\mu\nu\rho\sigma}
\tilde F_{\mu\nu}(x) \, \tilde F_{\rho\sigma}(x+\hat \mu+\hat \nu)
+\partial_\mu^\ast k^\prime_{\mu }(x),
\end{eqnarray}
where
\begin{equation}
k^\prime_{\mu }(x)=
k_{\mu }(x) + \sum_{\nu\rho\sigma} \gamma \epsilon_{\mu\nu\rho\sigma}
\frac{2\pi m_{\mu\nu}}{L^2} \left( F_{\rho\sigma}(x+\hat \mu+\hat \nu)
                                                   +F_{\rho\sigma}(x+\hat \mu) \right) .
\end{equation}
This result should be compared with eq.~(\ref{eq:chiral-anomaly-result}).
Indeed, it is possible to show that 
the coefficients of the polynomials of field tensor 
$\tilde F_{\mu\nu}(x)$ in eq.~(\ref{eq:chiral-anomaly-result}), 
$q_{[m]}(x)$, $\phi_{[m]\mu\nu}(x)$ and $\gamma_{[m,w]}$, 
are related to $\alpha$, $\beta_{\mu\nu}$ and $\gamma$ by the following bounds.

\vspace{1em}
\noindent{\bf Lemma \ref{sec:chiral-anomaly-in-finite-lattice}.3} \ \
\begin{eqnarray}
\label{eq:bound-q}
&& \hspace{-10mm} \left\vert q_{[m]}(x) - 
 \alpha -\beta_{\mu\nu} \frac{2\pi m_{\mu\nu}}{L^2}
-\gamma \epsilon_{\mu\nu\rho\sigma}
  \frac{(2\pi)^2 m_{\mu\nu} m_{\rho\sigma} }{L^4}
  \right\vert  \le 
\kappa_3 L^{\sigma_3} {\rm e}^{-L/2\rho},     \\
\label{eq:bound-phi}
&& \hspace{-10mm} \left\vert \phi_{[m]\mu\nu}(x) -\beta_{\mu\nu}- 2 \gamma 
\epsilon_{\mu\nu\rho\sigma}
  \frac{2\pi  m_{\rho\sigma} }{L^2}
  \right\vert  \le  
\kappa_3^\prime L^{\sigma_3} {\rm e}^{-L/2\rho},     \\
\label{eq:bound-gamma}
&& \hspace{-10mm} \left\vert \gamma_{[m,w]}- \gamma 
  \right\vert  \le  
\kappa_3^{\prime\prime} L^{\sigma_3} {\rm e}^{-L/2\rho}.
\end{eqnarray}

\vspace{1em}
\noindent  The proof of the bounds is given in the 
appendix~\ref{app:relation-to-infinite-lattice-phi}. 

\section{Exact cancellation of gauge anomaly}

We now consider the chiral anomaly which is given in terms of 
the overlap Dirac operator which satisfies the Ginsparg-Wilson relation as follows:
\begin{equation}
q(x)={\rm tr}\left\{
\gamma_5(1- D_L)(x,x) \right\},  
\end{equation}
where $D_L(x,y)$ is the finite-volume kernel of the Dirac operator. 
$q(x)$ is then defined for all admissible gauge fields 
and it is topological in the sense that
\begin{equation}
  \sum_{x \in \Gamma_4} q(x) = \text{integer}
\end{equation}
and also that 
\begin{equation}
\sum_{x \in \Gamma_4}  \delta q(x) = 0 , 
\end{equation}
for any local variation of the gauge field. 
This chiral anomaly  $q(x)$  can be  separated into two parts,  
the part defined in the infinite lattice and the part of finite-volume correction,
\begin{equation}
%\label{eq:q-separation}
q(x)=q_\infty(x) + \Delta q_\infty(x) . 
\end{equation}
Using eq.~(\ref{eq:finite-volume-kernel}), we explicitly have 
\begin{eqnarray}
q_\infty(x)&=&{\rm tr}\left\{
\gamma_5(1- D)(x,x) \right\}, \\
\Delta q_\infty (x)&=&\sum_{n \in \mathbb{Z}^4,n \not = 0}{\rm tr}\left\{
\gamma_5(1- D)(x,x+ n L) \right\}. 
\end{eqnarray} 
$q_\infty(x)$ is the chiral anomaly 
defined  in the infinite lattice with the periodic link variables. 
On the other hand, $\Delta_L q(x)$ satisfies the bound,
\begin{equation}
\vert \Delta q_\infty (x) \vert  \le \, \kappa \, {\rm e}^{- L/\varrho} . 
\end{equation}
Since $q_\infty(x)$ transforms as pseudo scalar, then 
we can show that the coefficients $\alpha$, $\beta_{\mu\nu}$
vanish identically. Moreover $\gamma$ can be evaluated 
in perturbation theory, giving the result 
$\gamma= \frac{1}{32\pi^2}$ \cite{Kikukawa:1998pd}.

Based on the result obtained in the previous subsections, 
we will now examine the cancellation of the gauge anomaly.
Let us consider an anomaly-free 
multiplet $ \{ \psi_L^\alpha(x) \, \vert \, \alpha=1,2,\cdots \}$
with the $U(1)$ charges satisfying the condition
\begin{equation}
  \sum_\alpha e_\alpha^3 = 0.
\end{equation}
In each contribution of a single Weyl fermion $\psi_L^\alpha(x)$ 
to the gauge anomaly, which is denoted by $q^\alpha(x)$, we scale 
the vector potential and the magnetic fluxes as 
\begin{equation}
  \tilde A_\mu(x) \rightarrow e_\alpha \tilde A_\mu(x), \qquad
  m_{\mu\nu} \rightarrow e_\alpha m_{\mu\nu},
\end{equation}
and consider the summation of them in the anomaly-free multiplet,
\begin{equation}
  \sum_\alpha e_\alpha \, q^\alpha(x) .
\end{equation}
In particular, we consider the summation of the first three terms
in the r.h.s. of eq.~(\ref{eq:chiral-anomaly-result}), 
\begin{eqnarray}
\sum_\alpha  e_\alpha A^\alpha(x)
=\sum_\alpha 
 \left\{ 
e_\alpha q^\alpha_{[m]}(x) %\frac{Q}{L^4} %
+ e_\alpha^2 \phi^\alpha_{[m]\mu\nu}(x) \tilde
F_{\mu\nu}(x)\right.\qquad \qquad \qquad \nonumber\\
\left.+ e_\alpha^3 \gamma^\alpha_{[m,w]} \epsilon_{\mu\nu\rho\sigma}
\tilde F_{\mu\nu}(x) \, \tilde F_{\rho\sigma}(x+\hat\mu+\hat\nu)
 \right\},\qquad
\end{eqnarray}
Then we can show the following lemma.

\vspace{1em}
\noindent{\bf Lemma \ref{sec:chiral-anomaly-in-finite-lattice}.4} \ \ 
For an anomaly-free multiplet, the summation of the gauge anomalies
$\sum_\alpha  e_\alpha A^\alpha(x)$ satisfies the bound
\begin{equation}
\label{eq:non-trivail-part-bound}
  \left\vert   \sum_\alpha  e_\alpha A^\alpha(x)
 \right\vert
\le \kappa_4 L^{\sigma_4} {\rm e}^{-L/2\rho},
\end{equation}
and can be written as the total-divergence of a gauge-invariant local
current,
\begin{equation}
\label{eq:non-trivail-part-k}
\sum_\alpha  e_\alpha A^\alpha(x) = \partial^\ast_\mu \Delta \tilde k_\mu(x).
\end{equation}

\vspace{1em}
\noindent{Proof:} \ \
From the bounds (\ref{eq:bound-q}), 
(\ref{eq:bound-phi}) and (\ref{eq:bound-gamma}), 
$e_\alpha A^\alpha(x)$ scales as $e_\alpha^3$ up to 
corrections of order ${\cal O}( L^\sigma {\rm e}^{-L/2\varrho})$.
The terms which scale as $e_\alpha^3$ cancel each other due to 
the anomaly-free condition and the remaining term satisfies
the bound eq.~(\ref{eq:non-trivail-part-bound}). 
Since $\sum_\alpha  e_\alpha Q^\alpha = 0$ where
$Q^\alpha=\sum_{x \in \Gamma_4} q^\alpha(x) = 
e_\alpha^2 \gamma \epsilon_{\mu\nu\rho\sigma}
(2\pi)^2 m_{\mu\nu} m_{\rho\sigma}$, 
 $\sum_\alpha  e_\alpha A^\alpha(x)$ satisfies
\begin{equation}
  \sum_{x \in \Gamma_4} 
  \left\{ \sum_\alpha  e_\alpha A^\alpha(x) \right\}= 0.
\end{equation}
Then we may apply the lemma \ref{sec:poincare-lemma-on-finite-lattice}b 
with an arbitrary reference point
$x_0$ to obtain eq.~(\ref{eq:non-trivail-part-k}).
This proves the lemma.

By the combination of lemma \ref{sec:chiral-anomaly-in-finite-lattice}.1, 
\ref{sec:chiral-anomaly-in-finite-lattice}.2 and 
\ref{sec:chiral-anomaly-in-finite-lattice}.4, we finally showed
that the gauge anomaly is cohomologically trivial. Namely, 
\begin{equation}
\label{eq:current-k}
  \sum_\alpha e_\alpha \, q^\alpha(x)
  = \partial^\ast_\mu  k_\mu(x),
\end{equation}
where $k_\mu(x)$ is a gauge-invariant local current defined by
\begin{equation}
   k_\mu(x) = \tilde k_\mu(x) + \Delta \tilde k_\mu(x).
\end{equation}
Thus we have shown that 
the current $k_\mu(x)$ which gives the 
the cohomologically trivial part of the chiral anomaly 
can be obtained  directly from the quantities calculable on the finite lattice. 
It is in sharp contrast with the current $k_\mu(x)$ given by
eq.~(\ref{eq:current-k-infinite-lattice})
which has been obtained throught the cohomological analysis in 
the infinite lattice.  
This is the  main result of our analysis.

\section{Discussion: Weyl fermion measure on the finite lattice}
\label{sec:shortcut-measure-term}

Finally we will argue how to construct the measure term\cite{Luscher:1998du}
directly from the cohomologically trivial part of the chiral 
anomaly on the finite lattice.

In the topological sector with the magnetic flux $m_{\mu\nu}$,
we choose an  one-parameter family of the admissible
U(1) gauge fields as 
\begin{equation}
  U^t(x,\mu)= {\rm e}^{i t \tilde A_\mu(x)} \times
  V_{[m]}(x,\mu),\qquad   0 \le t \le 1 .
\end{equation}
Let $\eta_\mu(x)$ be a real periodic vector field as a variational parameter.
Then we consider the following linear functional on the finite
lattice\cite{Luscher:1998du,Suzuki:1999qw}:
\begin{eqnarray}
\label{eq:measure-term-finite-volume}
\mathfrak{L}_\eta &=& i \int_0^1 dt \, 
{\rm Tr}_L \left\{ \hat P_L [ \partial_t \hat P_L, 
\delta_\eta \hat P_L ] \right\} 
\nonumber\\
&& +\int_0^1 dt \, \sum_{x \in \Gamma_4} %\mathbb{Z}^4}
\left\{\eta_\mu(x) \bar k_\mu(x)  + \tilde A_\mu(x) \delta_\eta 
\bar k_\mu(x) \right\},
\end{eqnarray}
where $\bar k_\mu(x)$ is a gauge-invariant local current,
which transforms as an axial vector field under the lattice symmetries
and which satisfies 
$\partial_\mu^\ast \bar k_\mu(x) = \sum_\alpha e_\alpha q^\alpha(x)$.
Such a current can be constructed from the current $k_\mu(x)$
in eq.~(\ref{eq:current-k}) by averaging over the lattice symmetries,
with the appropriate weights so at to project to the axial current
component.
This linear functional has the same form as $\mathfrak{L}^\star_\eta$ 
in \cite{Luscher:1998du}, the measure term in 
the infinite volume.
We can show that $\mathfrak{L}_\eta$ defined by 
eq.~(\ref{eq:measure-term-finite-volume}) satisfies all the 
properties required for the measure term {\it on the finite lattice}.

\vspace{1em}
\noindent{\bf Lemma \ref{sec:shortcut-measure-term}} 
The linear functional $\mathfrak{L}_\eta = \sum_{x\in \Gamma_4}
\eta_\mu(x) j_\mu(x)$ defined above has the following 
properties.

\begin{enumerate}
\item $j_\mu(x)$ is a local current, which is 
defined for all admissible gauge fields
and depends smoothly on the link variables.

\item $j_\mu(x)$ is gauge-invariant and transforms as an axial vector
  current under the lattice symmetries. 

\item The linear functional $\mathfrak{L}_\eta$
is a solution of the integrability condition
\begin{equation}
\delta_\eta \mathfrak{L}_\zeta - \delta_\zeta \mathfrak{L}_\eta
= i {\rm Tr}_L \left\{ \hat P_L [ \delta_\eta \hat P_L, \delta_\zeta
\hat P_L ] \right\}
\end{equation}
for all periodic variations $\eta_\mu(x)$ and $\zeta_\mu(x)$.

\item The anomalous conservation law 
$\partial_\mu^\ast j_\mu(x) = \sum_\alpha e_\alpha q^\alpha(x)$
holds.

\end{enumerate}

\noindent
The proof of the lemma goes just like the proof of 
the theorem 5.3 in \cite{Luscher:1998du} (section 6) and
we omit it here. A non-trivial point is the 
smoothness with respect to the link variables. This is because
the periodic vector potentials $\tilde A_\mu(x)$ do not
cover smoothly the space of the admissible gauge fields and 
at the singular points, 
the vector potentials can differ by 
the gauge functions $\omega(x)$
that are bounded polynomially in the lattice size $L$.
However,
$\mathfrak{L}_\eta$ is invariant under
gauge transformations $\tilde A_\mu(x) \rightarrow
\tilde A_\mu(x) + \partial_\mu \omega(x)$, 
for arbitrary gauge functions $\omega(x)$
that are bounded polynomially in the lattice size $L$.
Then it can be regarded as a smooth function of the 
link variables.

The global integrability condition
can be also established following the proof of the theorem 5.1 in 
\cite{Luscher:1998du} (section 10). 
Therefore the linear functional $\mathfrak{L}_\eta = \sum_{x\in \Gamma_4}
\eta_\mu(x) j_\mu(x)$ defined above provides a solution to
the measure term on the finite lattice.

\bigskip

\acknowledgments

The authors would like to thank H.~Suzuki for valuable discussions.
They are also grateful to M.~L\"uscher for useful comments.
D.K. is supported in part by the Japan
Society for Promotion of Science under the Predoctoral
Research Program No.~15-887.
Y.K. is supported in part by Grant-in-Aid 
for Scientific Research No.~14046207.

\appendix
\section{The bound on $\bm{ \| g \|}$ } 
\label{app:bound-on-norm-of-forms}

In this appendix, we give a proof of the bound eq.~(\ref{eq:bound-f-g}). 
For simplicity,  we set $x_0=0$. We prove the bound by induction. 
For $n=1$, the $0$-form $g$ vanishes identically
and the bound is satisfied trivially.  The $1$-form $g$ is given by
\begin{equation}
g(x)= \sum_{y_1=-L/2}^{x_1} f(y) \, dx_1
      = \left\{ 
      \begin{array}{cc}
    \displaystyle  - \sum_{y_1=x_1+1}^{L/2-1} f(y) \, dx_1 & \quad ( x_1 \ge 0 )  \\
     \displaystyle    \sum_{y_1=-L/2}^{x_1} f(y) \, dx_1 &  \quad ( x_1 < 0 ) 
      \end{array}
      \right. .
\end{equation}
Then, using $\vert f(x) \vert \le \| f \|_{p,\varrho}  {(1+\| x\|^{p} ) \,
   {\rm e}^{-\| x\|/\varrho} }$, 
we can estimate the absolute value of the coefficient of $g$ as 
\begin{eqnarray}
\vert g_1 (x) \vert 
& \le &
\left\{ 
      \begin{array}{cc}
    \displaystyle   
     \| f \|_{p,\varrho} \sum_{y_1=x_1+1}^{L/2-1}  {(1+\| y\|^{p} ) \,  {\rm e}^{-\| y\|/\varrho} }
    &  \quad ( x_1 \ge 0 )  \\
     \displaystyle   
     \| f \|_{p,\varrho} \sum_{y_1=-L/2}^{x_1}  {(1+\| y\|^{p} ) \,   {\rm e}^{-\| y\|/\varrho} }
     & \quad ( x_1 < 0 ) 
      \end{array}
      \right.
      \nonumber\\
& \le &   \| f \|_{p,\varrho}  \sum_{k=\|x\|}^\infty  {(1+k^{p} ) \, {\rm e}^{-k/\varrho} }
\nonumber\\
& \le &
\| f \|_{p,\varrho}  \, C_{p,\rho} {(1+\| x\|^{p} ) \,   {\rm e}^{-\| x\|/\varrho} }, 
\end{eqnarray}
where 
$C_{p,\rho} = \left[ 
\sum_{l=0}^p {}_pC_l \frac{\partial^l}{\partial \xi^l} \frac{1}{(1-{\rm e}^\xi)}
\right]_{\xi =-1/\varrho}$
which is independent of $L$.  Thus we have the bound on the norm of $g$ as
\begin{equation}
\label{app:bound-on-g-n=1}
\|g \|_{p,\varrho} \le C_{p,\varrho} \, \|f \|_{p,\varrho} . 
\end{equation}

Next we assume the bound eq.~(\ref{app:bound-on-g-n=1}) holds true  for $n-1$-dimensions $(n > 1)$ and consider the case of $n$-dimensions.  
For the closed $k+1$-form $f= u \, dx_n + v$, 
the $k$-form $g$ is given by the formula
\begin{equation}
g(x) = \delta_{x_n,0} \,  \bar g(x) + h(x)
+ \bar e(x) \, dx_n , 
\end{equation}
where $\bar g$ is obtained as 
\begin{equation}
\bar v = \bar d^\ast \bar g + \Delta \bar v; \qquad 
\bar v(x) \equiv \sum_{y_n=-L/2}^{L/2-1} v(y), 
\end{equation}
$h$ is defined by 
\begin{equation}
h(x) = (-1)^{k} \sum_{y_n=-L/2}^{x_n}
\left\{ v(y) - \delta_{y_n,0} \, \bar v \right\} \, d x_n 
=  \left\{  
\begin{array}{ll}
\displaystyle (-1)^{k+1}  \sum^{L/2-1}_{y_n=x_n+1}v(y) \, d x_n\qquad (x_n \ge 0) \\
\displaystyle  (-1)^{k} \sum^{x_n}_{y_n=-L/2}v(y)  \, d x_n  \qquad   (x_n < 0)
\end{array}
                 \right., 
\end{equation}
and $\bar e$ is obtained as
\begin{equation}
\bar u(x) = \bar d^\ast \bar e(x) + \Delta \bar u(x) ; \qquad 
\bar u(x) \equiv u(x)\vert_{x_n=L/2-1} . 
\end{equation}
Then the norm of $g$ is bounded by the sum of the norms of $\bar g$, 
$h$ and $\bar e$ and therefore we need to evaluate these three norms. 

The absolute value of the coefficient of $h$ can be estimated in just the
same manner as the case of $g$ for $n=1$:
\begin{eqnarray}
\vert h_{\mu_1\ldots\mu_{k+1}}(x)\vert
&\le& 
\left\{ 
\begin{array}{cc}
\displaystyle \|f\|_{p,\varrho} 
\sum^{L/2-1}_{y_n=x_n+1}  {(1+\| y\|^{p} ) \,   {\rm e}^{-\| y\|/\varrho} }
& \qquad (x_n \ge 0 ) \\
\displaystyle \|f\|_{p,\varrho} 
\sum^{x_n}_{y_n=-L/2}    {(1+\| y\|^{p} ) \,   {\rm e}^{-\| y\|/\varrho} }
& \qquad (x_n < 0) 
\end{array}
\right.
\nonumber\\
& \le &   \| f \|_{p,\varrho}  \sum_{k=\|x\|}^\infty  {(1+k^{p} ) \, {\rm e}^{-k/\varrho} }
\nonumber\\
& \le &
\| f \|_{p,\varrho}  \, C_{p,\rho} {(1+\| x\|^{p} ) \,   {\rm e}^{-\| x\|/\varrho} }.  
\end{eqnarray}
Thus we have the bound on the norm of $h$ as
\begin{equation}
\|h \|_{p,\varrho} \le C_{p,\varrho} \, \|f \|_{p,\varrho} . 
\end{equation}
As to the absolute value of $\bar v$, we have an estimate, 
\begin{eqnarray}
\vert \bar v_{\mu_1\ldots\mu_k}(x) \vert
&\le&   \|f \|_{p,\varrho} \, \sum_{y_n=-\infty}^\infty 
 {(1+\| y\|^{p} ) \,   {\rm e}^{-\| y\|/\varrho} } \nonumber\\
&\le& \|f \|_{p,\varrho} \, 
2 \, C_{p,\varrho} {(1+\| x\|^{p} ) \,   {\rm e}^{-\| x\|/\varrho} }, 
\end{eqnarray}
which implies the bound $ \| \bar v \| \le 2 C_{p,\varrho}  \|f \|_{p,\varrho}$.
Then, by the induction hypothesis, we have the bound on the 
norm of $\bar g$ as
\begin{equation}
 \| \bar g  \| \le D_{p,\varrho}  \|f \|_{p,\varrho} 
\end{equation}
for a constant $D_{p,\varrho}$ independent of $L$. 
As to $\bar u$, its norm is bounded by the norm of $f$, and 
by the induction hypothesis, we have the bound on the 
norm of $\bar e$ 
\begin{equation}
 \| \bar e  \| \le E_{p,\varrho}  \|f \|_{p,\varrho} 
\end{equation}
for a constant $E_{p,\varrho}$ independent of $L$.
Combining these three bounds, 
we finally obtain the bound on the norm of $g$, 
\begin{equation}
\|g\|_{p,\varrho}\le \bar C_{p,\varrho} \,\|f\|_{\,p,\varrho} , 
\end{equation}
where $\bar C_{p,\varrho}=C_{p,\varrho}+D_{p,\varrho}+E_{p,\varrho}$ 
is a constant which may depend on $p$ and $\varrho$, but 
not on the lattice size $L$. This completes the proof. 

\section{Relation between $\bm{g}$ and $\bm{g_\infty}$}
\label{app:relation-to-infinite-lattice}

We consider a closed form $f$ which is a local composite field of the gauge field
in the sense specified in 
sec.~\ref{sec:chiral-anomaly-in-finite-lattice-and-locality} with a 
reference point $x_0$.
From its locality property,  the norm of $f$ is bounded by a constant
$C_1$ independent of $L$ for the fixed $x_0$, $p$ and $\rho$, 
\begin{equation}
\| f \|_{x_0,p,\rho} \le C_1 . 
\end{equation}
Then lemma \ref{sec:poincare-lemma-on-finite-lattice}a asserts that
there exist two forms $g$ and
 $\Delta f$ which satisfies
 \begin{eqnarray}
&&  f(x) = d^\ast g(x) + \Delta f(x),  \\
&&  \| g \|_{x_0,p,\rho} \le C_2, \qquad 
   \vert \Delta f_{\mu_1 \ldots \mu_k} (x)
    \vert  \le  C_3  L^{p} {\rm e}^{-L/2\varrho} .
 \end{eqnarray}

On the other hand, $f$ can be expressed as 
\begin{equation}
f(x) = f_\infty(x) + \Delta f_\infty(x) , 
\end{equation}
where $f_\infty$ is a local form defined in the infinite lattice which satisfies
\begin{equation}
\vert f_{\mu_1 \ldots \mu_k \infty}(x) \vert 
\le c_1 (1+\| x-x_0 \|^p ) \, {\rm e}^{- \| x-x_0 \| / \varrho}, 
\end{equation}
and $\Delta f_\infty$ is a finite volume correction which satisfies the bound
\begin{equation}
\vert \Delta f_{\mu_1 \ldots \mu_k \infty}(x) \vert
\le c_2 \,  L^p \,  {\rm e}^{- L/ \varrho} . 
\end{equation}
If $f_\infty$ is also a closed form, 
\begin{equation}
d^\ast f_\infty(x) = 0 , 
\end{equation}
then the original Poincar\'e lemma \cite{Luscher:1998kn} asserts that 
there exists a form $g_\infty$ such that
\begin{equation}
f_\infty (x) = d^\ast g_\infty(x).  
\end{equation}
Then the question we address in this appendix is the relation between 
$g$ and $g_\infty$ restricted with the periodic gauge fields. 
We can show that the difference between these two forms
is a finite-volume correction which is suppressed exponentially in $L$. 
Namely, we have

\vspace{1em}
\noindent{\bf Lemma \ref{app:relation-to-infinite-lattice}} \ \
\begin{eqnarray}
\label{eq:relation-g-g-infty}
&& g(x) = g_\infty(x) + \Delta g(x) \qquad ( x \in \Gamma_{n[x_0]})  \\
&& \vert \Delta g(x) \vert \le c_3 L^{p} {\rm e}^{-L/2\rho} .
\end{eqnarray}

\vspace{1em}
\noindent{Proof:} \ \    
The proof of \ref{sec:poincare-lemma-on-finite-lattice}a,
\ref{sec:poincare-lemma-on-finite-lattice}b
holds true even if the lattice size is taken to be infinity $L\rightarrow \infty$, 
because of the locality property of the class of forms $f$ in consideration, and
it actually reduces in this limit to 
the proof of  the original Poincar\'e lemma in \cite{Luscher:1998kn}. Then
the constructed
$g$ simply reduces to $g_\infty$. 
Therefore, it suffices to 
compare these two solutions at a finite $L$ and in the limit $L \rightarrow \infty$ 
in each step of the induction of the proof of the lemma. 

For the case $n=1$, 
$0$-form $f$ can be expressed as an exterior
differential of $g$ which is defined by     
\begin{eqnarray}
g(x) &=& \sum^{x_1}_{y_1={x_0}_1-L/2} f(y) \, \, d x_1.
\end{eqnarray}
But  we can rewrite it as 
\begin{eqnarray}
g(x)&=&\sum^{x_1}_{y_1={x_0}_1-L/2} (  f_\infty (y) + \Delta f_\infty(y) )\, \, d x_1 
\nonumber\\
&=& g_\infty (x)- \sum^{{x_0}_1-L/2-1}_{y_1=-\infty} f_\infty(y) \, \, d x_1\,+
 \sum^{x_1}_{y_1={x_0}_1-L/2} \Delta f_\infty(y) \, \, d x_1 \nonumber\\
&=&g_\infty(x)+\Delta g(x) ,  
\end{eqnarray}
where $|\Delta g_1(x)| \le \kappa  L^{p} e^{-L/2\varrho}$ . 
The bound on $\Delta g$ follows from the locality property of $f_\infty$ 
and $\Delta f_\infty$. 
When $f$ is $1$-form, $g=g_\infty=0$. Thus the statement is proved for $n=1$.       

We next consider the case $n > 1$.  In this case, for a closed form $f= u dx_n + v $, 
$g$ is constructed from $v$ and two closed forms in $n-1$ dimensions defined by
\begin{eqnarray}
&& \bar v(x)\equiv \sum_{y_n=-L/2}^{L/2-1} v(y), \quad (x_1,\ldots,x_{n-1},y_n), 
\nonumber\\
&&\bar u(x)\equiv u(x)\vert_{x_n=x_{0n}+L/2-1}. 
\end{eqnarray}
$\bar v$ and $\bar u$ lead to the forms
$\bar g$ and $\bar e$, respectively, as
\begin{eqnarray}
&&\bar v = \bar d^\ast \bar g + \Delta \bar v ,   \nonumber\\
&&\bar u = \bar d^\ast \bar e + \Delta \bar u .  
\end{eqnarray}
Then $g$ is given by 
\begin{equation}
g(x) = \delta_{x_n,{x_0}_n} \bar g(x) + h(x)
+ \bar e(x) \, dx_n,
\end{equation}
where  
\begin{eqnarray}
&& h(x) \equiv (-1)^k \sum_{y_n=x_{0n}-L/2} ^{x_n}
\{ v(y)- \delta_{y_n,x_{0n}} \bar v \} dx_n . 
\end{eqnarray}
Since it holds true that 
\begin{eqnarray}
\bar v(x) &=& \sum_{y_n=-L/2}^{L/2-1} \{ v_\infty(y)+ \Delta v_\infty(y) \} \nonumber\\
              &=& \bar v_\infty(x)  
              -\sum_{y_n=-\infty }^{-L/2-1} v_\infty(y)
              -\sum_{y_n=L/2}^{\infty }  v_\infty(y)
              + \sum_{y_n=-L/2}^{L/2-1}  \Delta v_\infty(y) , \nonumber\\
\bar u(x) &=& \{ u_\infty(x) + \Delta u_\infty(x)\} \vert_{x_n=x_{0n}+L/2-1}, 
\end{eqnarray}
we have 
\begin{eqnarray}
\bar v(x) &=&  \bar v_\infty(x) + \Delta \bar v(x) ; \qquad 
         |\Delta \bar v_{\mu_1\ldots\mu_k} (x)| \le \kappa_1  L^{p} e^{-L/2\varrho}, \nonumber\\
\bar u(x) &=& \bar u_\infty(x) + \Delta \bar u(x) ; \qquad 
\bar u_\infty(x)=0 , \quad 
 |\Delta \bar u_{\mu_1\ldots\mu_{k-1}} (x)| \le \kappa_2  L^{p} e^{-L/2\varrho}. 
\end{eqnarray}
Then, by the induction hypothesis, we can infer that
\begin{eqnarray}
&& \bar g(x) = \bar g_\infty(x) + \Delta \bar g(x) \ ; \quad
|\Delta \bar g_{\mu_1\ldots\mu_{k+1}}(x)| \le \kappa_1^\prime  L^{p} e^{-L/2\varrho}, 
\nonumber\\
&& \bar e(x) = \bar e_\infty(x) + \Delta \bar e(x) ; \quad  \bar e_\infty(x) = 0 , \quad
|\Delta \bar e_{\mu_1\ldots\mu_{k}}(x)| \le \kappa_2^\prime  L^{p} e^{-L/2\varrho}. 
\end{eqnarray}
Also we have
\begin{eqnarray}
h(x) &=& 
 (-1)^k \sum_{y_n=x_{0n}-L/2}^{x_n} \{ v_\infty(y)- \delta_{y_n,x_{0n}} \bar v_\infty \} dx_n 
\nonumber\\
&+&   (-1)^k \sum_{y_n=x_{0n}-L/2}^{x_n} \{ \Delta v_\infty(y)- \delta_{y_n,x_{0n}} 
         \Delta \bar v_\infty \} dx_n \nonumber\\
&=& h_\infty(x) 
- (-1)^k \sum_{y_n=-\infty}^{x_{0n}-L/2-1} \{ v_\infty(y)- \delta_{y_n,x_{0n}}
 \bar v_\infty \} dx_n 
\nonumber\\
&& \qquad +  (-1)^k \sum_{y_n=x_{0n}-L/2}^{x_n} \{ \Delta v_\infty(y)- \delta_{y_n,x_{0n}} 
         \Delta \bar v_\infty \} dx_n \nonumber\\
         &=& h_\infty(x) + \Delta h(x) \ ; \qquad
          |\Delta h_{\mu_1\ldots\mu_{k+1}}(x)| \le \kappa_3  L^{p} e^{-L/2\varrho} . 
\end{eqnarray}
Then we can infer 
\begin{equation}
 g(x) = g_\infty(x) + \Delta g(x),  
\end{equation}
where 
\begin{equation}
\Delta g(x)= \delta_{x_n,{x_0}_n} \Delta \bar g(x) + \Delta h(x)
+ \Delta \bar e(x) \, dx_n \ ; 
\qquad 
 \vert \Delta g_{\mu_1\ldots\mu_{k+1}}(x) \vert \le \kappa_4 L^{p} {\rm e}^{-L/2\rho} .
\end{equation}
This result gives the proof of our statement.

\section{A proof of the bounds on $\bm{q_{[m]}(x)}$, 
$\bm{\phi_{[m]\mu\nu}(x)}$  and $\bm{\gamma_{[m,w]}}$ }
\label{app:relation-to-infinite-lattice-phi}

The first bound follows from the result 
eqs.~(\ref{eq:chiral-anomaly-result-infinite-lattice}) and 
(\ref{eq:current-k-infinite-lattice}), which is obtained through the 
cohomological analysis in the infinite lattice:  
for $U(x,\mu)=V_{[m]}(x,\mu)$, we have
\begin{eqnarray}
q_{[m]}(x) &=& q_\infty(x)\vert_{U=V_{[m]}} + \Delta q(x) \vert_{U=V_{[m]}} \nonumber\\
                 &=& 
                 \alpha +\beta_{\mu\nu} \frac{2\pi m_{\mu\nu}}{L^2}
                 +\gamma \epsilon_{\mu\nu\rho\sigma}
  \frac{(2\pi)^2 m_{\mu\nu} m_{\rho\sigma} }{L^4} 
  \nonumber\\
  && \quad + \partial_\mu^\ast 
  \{ k^{}_{\mu\infty}(x)\vert_{U=V_{[m]}}
+ \Delta k_{\mu\infty}(x) \vert_{U=V_{[m]}}  \} . 
\end{eqnarray}
In the second expression, the gauge-invariant local current 
$k_{\mu\infty}(x)\vert_{U={V_[m]}}$ should depend on
the gauge field through its field tensor which is now constant for $V_{[m]}(x,\mu)$. 
Then by the translational invariance the current should not depend on the site $x$
and its divergence vanishes identically. Then  the bound follows immediately.  

In order to show the second bound, 
we first recall that 
$\phi_{\mu\nu}(x)$ is calculated form the current $j_\mu(x,y)$ which is 
obtained by the first-derivative of the topological field $q(x)$
with respect to the vector potential $\tilde A_\mu(x)$. 
This current is evaluated for $V_{[m]}(x,\mu)$ as 
\begin{eqnarray}
j_\nu(x,y)\vert_{V_{[m]}} 
&=& \int_0^1 dt \left. \left(
\frac{\partial q(x)}{\partial 
\tilde A_\nu(y)} \right)_{\tilde A \rightarrow t \tilde A} \right\vert_{\tilde A =0}
\nonumber\\
&=&
 \left(\frac{\partial q_\infty(x)}{\partial \tilde A_\nu(y)} \right)_{\tilde A =0}
+ \left(\frac{\partial \Delta q_\infty(x)}{\partial \tilde A_\nu(y)} \right)_{\tilde A =0}
\nonumber\\
&=& 
 \sum_{n \in \mathbb{Z}^4}
  \left(\frac{\partial q_\infty(x)}{\partial  A_\nu(y+n L)} \right)_{U=V_{[m]}}
 + \left(\frac{\partial \Delta q_\infty(x)}{\partial \tilde A_\nu(y)} \right)_{\tilde A =0}
\qquad ( x,y \in \Gamma_4 ), 
\nonumber\\
\end{eqnarray}
where in the third line we have used eq.~(\ref{eq:q-separation})  and 
in the last line we have taken into account that $\tilde A_\mu(x)$ is a 
periodic field in the infinite lattice. $A_\mu(x)$ is the vector
potential introduced in \cite{Luscher:1998kn}. 

The term with $n=0$ in the first term of the r.h.s.,  
\begin{equation}
\left(\frac{\partial q_\infty(x)}{\partial  A_\nu(y)} \right) 
\qquad ( x,y \in \mathbb{Z}^4 ), 
\end{equation}
is now defined on the infinite lattice.  
This term is related to 
the corresponding current 
$j_{\mu\infty}(x,y)$ in the infinite lattice which appears
in the original cohomological analysis through the 
parameter integral,  
\begin{equation}
j_{\nu\infty}(x,y)
= \int_0^1 dt   
\left(\frac{\partial q_\infty(x)}{\partial  A_\nu(y)} 
\right)_{ A \rightarrow t A}. 
\end{equation}
From this current,  the tensor field $\phi_{\mu\nu\infty}(x,y)$ descends as  
\begin{equation}
j_{\nu\infty}(x,y) = \theta_{\nu\mu \infty}(x,y) \overleftarrow{\partial_\mu^\ast} , \qquad
\frac{1}{2}\sum_{z \in \mathbb{Z}_4}
\theta_{\mu\nu\infty}(z,x) = \phi_{\mu\nu\infty}(x) , 
\end{equation}
and it is evaluated  further as
\begin{equation}
\label{eq:second-step-result-infty}
 \phi_{\mu\nu\infty}(x) = \beta_{\mu\nu}+
\gamma \epsilon_{\mu\nu\rho\sigma}  F_{\rho\sigma}(x)
+\partial_\lambda^\ast  t_{\lambda\mu\nu\infty}(x).
\end{equation}
But we note here that in the above analysis we might have applied 
the Poincar\'e lemma
before doing the parameter integral. Namely, 
we have
\begin{equation}
 \left(\frac{\partial q_\infty(x)}{\partial  A_\nu(y)} \right)
= \check \theta_{\nu\mu \infty}(x,y) \overleftarrow{\partial_\mu^\ast} , \qquad
\frac{1}{2}\sum_{z \in \mathbb{Z}_4}
\check \theta_{\mu\nu\infty}(z,x) =\check \phi_{\mu\nu\infty}(x) 
\end{equation}
and  
\begin{equation}
 \check \phi_{\mu\nu\infty}(x) = \check \beta_{\mu\nu} 
+\check \gamma \epsilon_{\mu\nu\rho\sigma}  F_{\rho\sigma}(x)
+\partial_\lambda^\ast   \check t_{\lambda\mu\nu\infty}(x) . 
\end{equation}
Since 
the parameter integral of this expression 
should reproduce the above result eq.~(\ref{eq:second-step-result-infty}),  
$\check \beta_{\mu\nu}$, $\check \gamma$ and $\check  t_{\lambda\mu\nu\infty}(x)$ 
are related to $\beta_{\mu\nu}$, $\gamma$ and $ t_{\lambda\mu\nu\infty}(x)$, respectively as follows: 
\begin{equation}
\check \beta_{\mu\nu}= \beta_{\mu\nu}, 
\quad
\frac{1}{2} \check \gamma = \gamma, \quad
\int_0^1 dt   \left. \check t_{\lambda\mu\nu}(x) \right\vert_{A \rightarrow t A}
= t_{\lambda\mu\nu}(x). 
\end{equation}

The tensor field $\phi_{\mu\nu}(x)$ on the finite lattice, on the other hand, 
is obtained from $\theta_{\mu\nu}(x,y)$ as in eqs.~(\ref{eq:tilde-phi}), 
(\ref{eq:Delta-Phi}) and (\ref{eq:phi}). 
The difference between the solutions of these two lemma is, 
as shown in the appendix~\ref{app:relation-to-infinite-lattice},  
an exponentially small correction and therefore we can infer 
\begin{equation}
\theta_{\mu\nu}(x,y) \vert_{V_{[m]}}  = \check \theta_{\mu\nu\infty}(x,y) \vert_{V_{[m]}} 
+\Delta \theta_{\mu\nu}(x,y), \qquad 
\vert \Delta \theta_{\mu\nu}(x,y)\vert \le c_1
L^{\sigma_1} {\rm e}^{-L/2\rho}.  
\end{equation}
This immediately implies 
\begin{equation}
\phi_{[m]\mu\nu}(x)  = \check \phi_{\mu\nu\infty}(x) \vert_{V_{[m]}} 
+\Delta \phi_{\mu\nu}(x)\vert_{V_{[m]}}, \qquad 
\vert \Delta \phi_{\mu\nu}(x)\vert \le c_2
L^{\sigma_1} {\rm e}^{-L/2\rho},  
\end{equation}
while 
$\check \phi_{\nu\mu \infty}(x)$ evaluates for $V_{[m]}(x,\mu)$ as
\begin{equation}
 \check \phi_{\mu\nu\infty}(x) \vert_{V_{[m]}}= \beta_{\mu\nu}+
2 \gamma \epsilon_{\mu\nu\rho\sigma}  \frac{2\pi m_{\rho\sigma}}{L^2},
\end{equation}
because the gauge-invariant tensor field $\check t_{\lambda\mu\nu\infty}(x) $
is a constant for this case. 
This proves the second bound. 

As to the third bound, we first recall the fact that 
the dependence of $\gamma_{[m,w]}$ on the gauge potential $\tilde A_\mu(x)$
is almost excluded except the dependence on the Wilson line $w_\mu$. 
This dependence on $\tilde A_\mu(x)$ (or $w_\mu$)
is in fact exponentially small.  One way
to see this is to repeat the cohomological analysis for 
$\epsilon_{\rho\sigma\tau\lambda} \, \gamma_{[m,w]}(x)
=(1/2)\sum_{z \in \Gamma_4} \omega_{\rho\sigma\tau\lambda}(z,x)$
which satisfies 
$\partial_\rho^\ast \epsilon_{\rho\sigma\tau\lambda}\, \gamma_{[m,w]} (x) = 0$,
regarding it as a tensor field of rank four,  just like for $\phi_{\mu\nu}(x)$. 
Then we obtain an expression similar to
eq.~(\ref{eq:chiral-anomaly-phi-in-xi-check}). But in this case the tensor fields 
which correspond to 
$\check \xi$ and $\Delta \Xi$ in eq.~(\ref{eq:chiral-anomaly-phi-in-xi-check}) should 
be antisymmetric tensors of rank six
and should vanish identically in four dimensions. Therefore we have 
\begin{equation}
 \epsilon_{\rho\sigma\tau\lambda} \gamma_{[m,w]} 
=  \epsilon_{\rho\sigma\tau\lambda}  \gamma_{[m,w]}\vert_{\tilde A = 0}
+ \sum_{y \in \Gamma_4} \Delta j_{\rho\sigma\tau\lambda\mu}(x,y) 
\tilde A_\mu(y) ,
\quad 
\vert \Delta j_{\rho\sigma\tau\lambda\mu}(x,y)\vert \le c_3
L^{\sigma_3} {\rm e}^{-L/2\rho}
\end{equation}
and the difference $\gamma_{[m,w]} -\gamma_{[m,w]}\vert_{\tilde A = 0}$ 
is indeed exponentially small.

We next recall that $\gamma_{[m,w]}$ is calculated form 
the current $j_{\mu\nu\rho}(x,y)$ which is 
obtained by the first-derivative of the tensor field $\phi_{\mu\nu}(x)$
with respect to the vector potential $\tilde A_\mu(x)$.  This current is 
evaluated for $V_{[m]}(x,\mu)$ as 
\begin{eqnarray}
\label{eq:current-to-gamma-for-Vm}
\left. j_{\mu\nu\rho}(x,y) \right\vert_{V_{[m]}}
&=& \left. 
\int_0^1 dt \left(
\frac{\partial \phi_{\mu\nu}(x)}{\partial 
\tilde A_\rho(y)} \right)_{\tilde A \rightarrow t \tilde A} \right\vert_{\tilde A=0}
\nonumber\\
&=& 
 \left(
\frac{\partial \phi_{\mu\nu}(x)}{\partial 
\tilde A_\rho(y)} \right)_{\tilde A=0}. 
\end{eqnarray}
The tensor field $( \partial \phi_{\mu\nu}(x) / \partial \tilde A_\rho(y))_{\tilde A=0}$
in turn is calculated from the derivative of $j_{\mu}(x,z)$  with respect to the vector potential, that is the second  derivative of the topological field $q(x)$:
\begin{eqnarray}
\label{eq:current-to-phi-del-A}
\left( \frac{ \partial j_{\mu}(x,z)}{\partial \tilde A_\rho(y)} \right)_{\tilde A=0}
&=&
\left.  
\int_0^1 dt \, t \, \left(
\frac{\partial^2 q(x) }{\partial \tilde A_\rho(y) \partial \tilde A_\mu(z) } 
\right)_{\tilde A \rightarrow \tilde t A} \right\vert_{\tilde A=0}
\nonumber\\
&=&
\frac{1}{2} \left(
\frac{\partial^2 q(x) }{\partial \tilde A_\rho(y) \partial \tilde A_\mu(z) } 
\right)_{\tilde A=0} 
\nonumber\\
&=&
\sum_{m,n\in \mathbb{Z}^4}
\frac{1}{2} \left(
\frac{\partial^2 q_\infty (x) }{\partial  A_\rho(y+mL) \partial  A_\mu(z+nL) } 
\right)_{U=V_{[m]}}
+ \frac{1}{2} \left(
\frac{\partial^2 \Delta q_\infty(x) }{\partial \tilde A_\rho(y) \partial \tilde A_\mu(z) } 
\right)_{\tilde A=0} , 
\nonumber\\
&& \qquad\qquad\qquad\qquad\qquad\qquad\qquad\qquad
\qquad\qquad  (x,y,z \in \Gamma_4). 
\end{eqnarray}

The term with $m=n=0$ in the r.h.s. of eq.~(\ref{eq:current-to-phi-del-A})
\begin{equation}
\label{eq:term-with-m=n=0}
\left(
\frac{\partial^2 q_\infty (x) }{\partial  A_\rho(y) \partial  A_\mu(z) } 
\right)
\end{equation}
is again defined in the infinite lattice. 
It is first related to the differentiation of current $j_{\mu\infty}(x,y)$
with respect to the vector potential, by the parameter integral:
\begin{equation}
\left(\frac{\partial j_{\mu \infty}(x,z)}{\partial A_\rho(y)}\right)
= \int_0^1 dt \, t 
 \left(
\frac{\partial^2 q_\infty (x) }{\partial  A_\rho(y) \partial  A_\mu(z) } 
\right)_{A \rightarrow t A} . 
\end{equation}
From this current, $\left(\partial \phi_{\mu\nu \infty}(x) / \partial A_\rho(y) \right)$
is then obtained by the applications of the Poincar\'e and this in turn  is related to 
the current $j_{\mu\nu\rho \infty}(x,y)$ in the infinite lattice as 
\begin{equation}
 j_{\mu\nu\rho \infty}(x,y) =
\int_0^1 dt \left(
\frac{\partial \phi_{\mu\nu \infty}(x)}{\partial A_\rho(y)} \right)_{ A \rightarrow t  A}. 
\end{equation}

$j_{\mu\nu\rho \infty}(x,y)$ is evaluated 
by the cohomological analysis in the infinite lattice
as
\begin{eqnarray}
j_{\mu\nu\rho \infty}(x,y) &=& 
\xi_{\mu\nu\rho\sigma \infty}(x,y) \overleftarrow{\partial_\sigma^\ast}, \\
\xi_{\mu\nu\rho\sigma \infty}(x,y)&=& 
 2 \gamma \, \epsilon_{\mu\nu\rho\sigma}\delta_{x,y-\hat\mu-\hat\nu}
+\partial_\lambda^\ast
\kappa_{\lambda\mu\nu\rho\sigma \infty}(x,y)
+\theta_{\mu\nu\rho\sigma\tau \infty}(x,y) 
\overleftarrow{\partial_\tau^\ast} . 
\end{eqnarray}
But we note again that the same cohomological analysis may be applied 
to the current
$j^\prime_{\mu\nu\rho \infty}(x,y)$ which is obtained 
starting from eq.~(\ref{eq:term-with-m=n=0}), but 
without the parameter integrals. 
The result reads 
\begin{eqnarray}
j^\prime_{\mu\nu\rho \infty}(x,y) &=& 
\xi^\prime_{\mu\nu\rho\sigma \infty}(x,y) \overleftarrow{\partial_\sigma^\ast}, \\
\xi^\prime_{\mu\nu\rho\sigma \infty}(x,y)&=& 
 2 \gamma^\prime \, \epsilon_{\mu\nu\rho\sigma}\delta_{x,y-\hat\mu-\hat\nu}
+\partial_\lambda^\ast
\kappa^\prime_{\lambda\mu\nu\rho\sigma \infty}(x,y)
+\theta^\prime_{\mu\nu\rho\sigma\tau \infty}(x,y) 
\overleftarrow{\partial_\tau^\ast}, 
\end{eqnarray}
where $\gamma^\prime$, $\kappa^\prime$ and $\theta^\prime$ are 
related to their counterparts, 
$\gamma$, $\kappa$ and $\theta$,  as
\begin{eqnarray}
&&\gamma^\prime \, \int_0^1 ds\,  \int_0^1 dt \, t  =  \frac{1}{2} \gamma^\prime
= \gamma, \\
&& \int_0^1 ds\,  \int_0^1 dt \, t \, \, 
\kappa^\prime_{\lambda\mu\nu\rho\sigma \infty}(x,y)
\vert_{A \rightarrow st A} = \kappa_{\lambda\mu\nu\rho\sigma \infty}(x,y), \\
&&
\int_0^1 ds\,  \int_0^1 dt \, t \, \, 
\theta^\prime_{\mu\nu\rho\sigma\tau \infty}(x,y) 
\vert_{A \rightarrow st A} = \theta_{\mu\nu\rho\sigma\tau \infty}(x,y) . 
\end{eqnarray}

On the other hand, in obtaining $\gamma_{[m,w]}\vert_{\tilde A=0}$  
from eqs.~(\ref{eq:current-to-gamma-for-Vm}) and (\ref{eq:current-to-phi-del-A}), 
by the repeat applications of the Poincar\'e lemma, 
we may consider to apply 
the 
original Poincar\'e lemma in the infinite lattice  
to the term with $m=n=0$ 
in the r.h.s. of eq.~(\ref{eq:current-to-phi-del-A}), 
which is defined in the infinite lattice,  and  also to its descendants. 
The difference between the solutions of these two lemma is, 
as shown in the appendix~\ref{app:relation-to-infinite-lattice},  
an exponentially small correction and then we can infer 
\begin{equation}
\gamma_{[m,w]}\vert_{\tilde A=0} = \frac{1}{2} \gamma^\prime 
+ \Delta \gamma ; \qquad
\vert \Delta \gamma \vert
\le c_{4}
L^{\sigma_{2}} {\rm e}^{-L/2\rho} . 
\end{equation}
Taking into account that $\gamma^\prime = 2 \gamma$, we obtain
the third bound. 

%\listoftables           % ONLY IN DRAFT MODE
%\listoffigures          % ONLY IN DRAFT MODE


\begin{thebibliography}{999}
%\bibitem{proc} F. Nesti and P. Dall'Aglio, \emph{Sample proceedings in
%                JHEP format}, SISSA 2001.
%\bibitem{fltf}  Maths Dahlgren, {\it Package {\tt floatflt}, distributed 
%                with \LaTeXe{} 96/06/01}, 1994-1996.
%\bibitem{LC}    M. Goossens, F. Mittelbach, A. Samarin, 
%                {\it The \LaTeX{} Companion}, Addison-Wesley 1994.
%\bibitem{TeXbook} D. E. Knuth, {\it The \TeX book}, Addison-Wesley 1986.


%\cite{Ginsparg:1981bj}
\bibitem{Ginsparg:1981bj}
P.~H.~Ginsparg and K.~G.~Wilson,
%``A Remnant Of Chiral Symmetry On The Lattice,''
Phys.\ Rev.\ D {\bf 25}, 2649 (1982).
%%CITATION = PHRVA,D25,2649;%%


%\cite{Neuberger:1997fp}
\bibitem{Neuberger:1997fp}
H.~Neuberger,
%``Exactly massless quarks on the lattice,''
Phys.\ Lett.\ B {\bf 417}, 141 (1998)
[arXiv:hep-lat/9707022].
%%CITATION = HEP-LAT 9707022;%%

%\cite{Hasenfratz:1998ri}
\bibitem{Hasenfratz:1998ri}
P.~Hasenfratz, V.~Laliena and F.~Niedermayer,
%``The index theorem in QCD with a finite cut-off,''
Phys.\ Lett.\ B {\bf 427}, 125 (1998)
[arXiv:hep-lat/9801021].
%%CITATION = HEP-LAT 9801021;%%

%\cite{Neuberger:1998wv}
\bibitem{Neuberger:1998wv}
H.~Neuberger,
%``More about exactly massless quarks on the lattice,''
Phys.\ Lett.\ B {\bf 427}, 353 (1998)
[arXiv:hep-lat/9801031].
%%CITATION = HEP-LAT 9801031;%%

%\cite{Hasenfratz:1998jp}
\bibitem{Hasenfratz:1998jp}
P.~Hasenfratz,
%``Lattice QCD without tuning, mixing and current renormalization,''
Nucl.\ Phys.\ B {\bf 525}, 401 (1998)
[arXiv:hep-lat/9802007].
%%CITATION = HEP-LAT 9802007;%%


%\cite{Hernandez:1998et}
\bibitem{Hernandez:1998et}
P.~Hernandez, K.~Jansen and M.~L\"uscher,
%``Locality properties of Neuberger's lattice Dirac operator,''
Nucl.\ Phys.\ B {\bf 552}, 363 (1999)
[arXiv:hep-lat/9808010].
%%CITATION = HEP-LAT 9808010;%%

%


%\cite{Narayanan:wx}
\bibitem{Narayanan:wx}
R.~Narayanan and H.~Neuberger,
%``Infinitely Many Regulator Fields For Chiral Fermions,''
Phys.\ Lett.\ B {\bf 302}, 62 (1993)
[arXiv:hep-lat/9212019].
%%CITATION = HEP-LAT 9212019;%%
%

%\cite{Narayanan:sk}
\bibitem{Narayanan:sk}
R.~Narayanan and H.~Neuberger,
%``Chiral Determinant As An Overlap Of Two Vacua,''
Nucl.\ Phys.\ B {\bf 412}, 574 (1994)
[arXiv:hep-lat/9307006].
%%CITATION = HEP-LAT 9307006;%%

%\cite{Narayanan:ss}
\bibitem{Narayanan:ss}
R.~Narayanan and H.~Neuberger,
%``Chiral Fermions On The Lattice,''
Phys.\ Rev.\ Lett.\  {\bf 71}, 3251 (1993)
[arXiv:hep-lat/9308011].
%%CITATION = HEP-LAT 9308011;%%

%\cite{Narayanan:1994gw}
\bibitem{Narayanan:1994gw}
R.~Narayanan and H.~Neuberger,
%``A Construction of lattice chiral gauge theories,''
Nucl.\ Phys.\ B {\bf 443}, 305 (1995)
[arXiv:hep-th/9411108].
%%CITATION = HEP-TH 9411108;%%
%\cite{Narayanan:1993gq}

\bibitem{Narayanan:1993gq}
R.~Narayanan,
%``Recent developments in chiral gauge theories: Approach of infinitely many fermi fields,''
Nucl.\ Phys.\ Proc.\ Suppl.\  {\bf 34}, 95 (1994)
[arXiv:hep-lat/9311014].
%%CITATION = HEP-LAT 9311014;%%

%\cite{Neuberger:1999ry}
\bibitem{Neuberger:1999ry}
H.~Neuberger,
%``Chiral fermions on the lattice,''
Nucl.\ Phys.\ Proc.\ Suppl.\  {\bf 83}, 67 (2000)
[arXiv:hep-lat/9909042].
%%CITATION = HEP-LAT 9909042;%%

%\cite{Narayanan:1996cu}
\bibitem{Narayanan:1996cu}
R.~Narayanan and H.~Neuberger,
%``Anomaly free U(1) chiral gauge theories on a two dimensional torus,''
Nucl.\ Phys.\ B {\bf 477}, 521 (1996)
[arXiv:hep-th/9603204].
%%CITATION = HEP-TH 9603204;%%

%\cite{Huet:1996pw}
\bibitem{Huet:1996pw}
P.~Y.~Huet, R.~Narayanan and H.~Neuberger,
%``Overlap formulation of Majorana--Weyl fermions,''
Phys.\ Lett.\ B {\bf 380}, 291 (1996)
[arXiv:hep-th/9602176].
%%CITATION = HEP-TH 9602176;%%

%\cite{Narayanan:1997by}
\bibitem{Narayanan:1997by}
R.~Narayanan and J.~Nishimura,
%``Parity-invariant lattice regularization of a three-dimensional  gauge-fermion system,''
Nucl.\ Phys.\ B {\bf 508}, 371 (1997)
[arXiv:hep-th/9703109].
%%CITATION = HEP-TH 9703109;%%


%\cite{Kikukawa:1997qh}
\bibitem{Kikukawa:1997qh}
Y.~Kikukawa and H.~Neuberger,
%``Overlap in odd dimensions,''
Nucl.\ Phys.\ B {\bf 513}, 735 (1998)
[arXiv:hep-lat/9707016].
%%CITATION = HEP-LAT 9707016;%%

%\cite{Narayanan:1996kz}
\bibitem{Narayanan:1996kz}
R.~Narayanan and H.~Neuberger,
%``Massless composite fermions in two dimensions and the overlap,''
Phys.\ Lett.\ B {\bf 393}, 360 (1997)
[Phys.\ Lett.\ B {\bf 402}, 320 (1997)]
[arXiv:hep-lat/9609031].
%%CITATION = HEP-LAT 9609031;%%

%\cite{Kikukawa:1997md}
\bibitem{Kikukawa:1997md}
Y.~Kikukawa, R.~Narayanan and H.~Neuberger,
%``Finite size corrections in two dimensional gauge theories and a  quantitative chiral test of the overlap,''
Phys.\ Lett.\ B {\bf 399}, 105 (1997)
[arXiv:hep-th/9701007].
%%CITATION = HEP-TH 9701007;%%

%\cite{Kikukawa:1997dv}
\bibitem{Kikukawa:1997dv}
Y.~Kikukawa, R.~Narayanan and H.~Neuberger,
%``Monte Carlo evaluation of a fermion number violating observable in 2D,''
Phys.\ Rev.\ D {\bf 57}, 1233 (1998)
[arXiv:hep-lat/9705006].
%%CITATION = HEP-LAT 9705006;%%

%\cite{Kaplan:1992bt}
\bibitem{Kaplan:1992bt}
D.~B.~Kaplan,
%``A Method for simulating chiral fermions on the lattice,''
Phys.\ Lett.\ B {\bf 288}, 342 (1992)
[arXiv:hep-lat/9206013].
%%CITATION = HEP-LAT 9206013;%%

%\cite{Shamir:1993zy}
\bibitem{Shamir:1993zy}
Y.~Shamir,
%``Chiral fermions from lattice boundaries,''
Nucl.\ Phys.\ B {\bf 406}, 90 (1993)
[arXiv:hep-lat/9303005].
%%CITATION = HEP-LAT 9303005;%%

%\cite{Furman:ky}
\bibitem{Furman:ky}
V.~Furman and Y.~Shamir,
%``Axial Symmetries In Lattice QCD With Kaplan Fermions,''
Nucl.\ Phys.\ B {\bf 439}, 54 (1995)
[arXiv:hep-lat/9405004].
%%CITATION = HEP-LAT 9405004;%%

%\cite{Blum:1996jf}
\bibitem{Blum:1996jf}
T.~Blum and A.~Soni,
%``QCD with domain wall quarks,''
Phys.\ Rev.\ D {\bf 56}, 174 (1997)
[arXiv:hep-lat/9611030].
%%CITATION = HEP-LAT 9611030;%%

%\cite{Blum:1997mz}
\bibitem{Blum:1997mz}
T.~Blum and A.~Soni,
%``Domain wall quarks and kaon weak matrix elements,''
Phys.\ Rev.\ Lett.\  {\bf 79}, 3595 (1997)
[arXiv:hep-lat/9706023].
%%CITATION = HEP-LAT 9706023;%%

%\cite{Vranas:1997da}
\bibitem{Vranas:1997da}
P.~M.~Vranas,
%``Chiral symmetry restoration in the Schwinger model with domain wall  fermions,''
Phys.\ Rev.\ D {\bf 57}, 1415 (1998)
[arXiv:hep-lat/9705023].
%%CITATION = HEP-LAT 9705023;%%

%\cite{Neuberger:1997bg}
\bibitem{Neuberger:1997bg}
H.~Neuberger,
%``Vector like gauge theories with almost massless fermions on the  lattice,''
Phys.\ Rev.\ D {\bf 57}, 5417 (1998)
[arXiv:hep-lat/9710089].
%%CITATION = HEP-LAT 9710089;%%

%\cite{Kikukawa:1999sy}
\bibitem{Kikukawa:1999sy}
Y.~Kikukawa and T.~Noguchi,
%``Low energy effective action of domain-wall fermion and the  Ginsparg-Wilson relation,''
arXiv:hep-lat/9902022.
%%CITATION = HEP-LAT 9902022;%%

%% end for overlap 

%\cite{Luscher:1998pq}
\bibitem{Luscher:1998pq}
M.~L\"uscher,
%``Exact chiral symmetry on the lattice and the Ginsparg-Wilson relation,''
Phys.\ Lett.\ B {\bf 428}, 342 (1998)
[arXiv:hep-lat/9802011].
%%CITATION = HEP-LAT 9802011;%%

%\cite{Luscher:1998kn}
\bibitem{Luscher:1998kn}
M.~L\"uscher,
%``Topology and the axial anomaly in abelian lattice gauge theories,''
Nucl.\ Phys.\ B {\bf 538}, 515 (1999)
[arXiv:hep-lat/9808021].
%%CITATION = HEP-LAT 9808021;%%

%\cite{Luscher:1998du}
\bibitem{Luscher:1998du}
M.~L\"uscher,
%``Abelian chiral gauge theories on the lattice with exact gauge  invariance,''
Nucl.\ Phys.\ B {\bf 549}, 295 (1999)
[arXiv:hep-lat/9811032].
%%CITATION = HEP-LAT 9811032;%%

%\cite{Luscher:1999un}
\bibitem{Luscher:1999un}
M.~L\"uscher,
%``Weyl fermions on the lattice and the non-abelian gauge anomaly,''
Nucl.\ Phys.\ B {\bf 568}, 162 (2000)
[arXiv:hep-lat/9904009].
%%CITATION = HEP-LAT 9904009;%%

%\cite{Luscher:1999mt}
\bibitem{Luscher:1999mt}
M.~L\"uscher,
%``Chiral gauge theories on the lattice with exact gauge invariance,''
Nucl.\ Phys.\ Proc.\ Suppl.\  {\bf 83}, 34 (2000)
[arXiv:hep-lat/9909150].
%%CITATION = HEP-LAT 9909150;%%

%\cite{Luscher:2000hn}
\bibitem{Luscher:2000hn}
M.~L\"uscher,
%``Chiral gauge theories revisited,''
arXiv:hep-th/0102028.
%%CITATION = HEP-TH 0102028;%%

%\cite{Suzuki:1999qw}
\bibitem{Suzuki:1999qw}
H.~Suzuki,
%``Gauge invariant effective action in Abelian chiral gauge theory on the  lattice,''
Prog.\ Theor.\ Phys.\  {\bf 101}, 1147 (1999)
[arXiv:hep-lat/9901012].
%%CITATION = HEP-LAT 9901012;%%

%\cite{Neuberger:2000wq}
\bibitem{Neuberger:2000wq}
H.~Neuberger,
%``Noncompact chiral U(1) gauge theories on the lattice,''
Phys.\ Rev.\ D {\bf 63}, 014503 (2001)
[arXiv:hep-lat/0002032].
%%CITATION = HEP-LAT 0002032;%%

%\cite{Fujiwara:1999fi}
\bibitem{Fujiwara:1999fi}
T.~Fujiwara, H.~Suzuki and K.~Wu,
%``Non-commutative differential calculus and the axial anomaly in Abelian  lattice gauge theories,''
Nucl.\ Phys.\ B {\bf 569}, 643 (2000)
[arXiv:hep-lat/9906015].
%%CITATION = HEP-LAT 9906015;%%

%\cite{Fujiwara:1999fj}
\bibitem{Fujiwara:1999fj}
T.~Fujiwara, H.~Suzuki and K.~Wu,
%``Axial anomaly in lattice Abelian gauge theory in arbitrary dimensions,''
Phys.\ Lett.\ B {\bf 463}, 63 (1999)
[arXiv:hep-lat/9906016].
%%CITATION = HEP-LAT 9906016;%%

%\cite{Kikukawa:1998pd}
\bibitem{Kikukawa:1998pd}
Y.~Kikukawa and A.~Yamada,
%``Weak coupling expansion of massless QCD with a Ginsparg-Wilson fermion  and axial U(1) anomaly,''
Phys.\ Lett.\ B {\bf 448}, 265 (1999)
[arXiv:hep-lat/9806013].
%%CITATION = HEP-LAT 9806013;%%

%\cite{Fujikawa:1998if}
\bibitem{Fujikawa:1998if}
K.~Fujikawa,
%``A continuum limit of the chiral Jacobian in lattice gauge theory,''
Nucl.\ Phys.\ B {\bf 546}, 480 (1999)
[arXiv:hep-th/9811235].
%%CITATION = HEP-TH 9811235;%%

%\cite{Adams:1998eg}
\bibitem{Adams:1998eg}
D.~H.~Adams,
%``Axial anomaly and topological charge in lattice gauge theory with  overlap-Dirac operator,''
Annals Phys.\  {\bf 296}, 131 (2002)
[arXiv:hep-lat/9812003].
%%CITATION = HEP-LAT 9812003;%%

%\cite{Suzuki:1998yz}
\bibitem{Suzuki:1998yz}
H.~Suzuki,
%``Simple evaluation of chiral Jacobian with the overlap Dirac operator,''
Prog.\ Theor.\ Phys.\  {\bf 102}, 141 (1999)
[arXiv:hep-th/9812019].
%%CITATION = HEP-TH 9812019;%%

%\cite{Chiu:1998xf}
\bibitem{Chiu:1998xf}
T.~W.~Chiu,
%``The axial anomaly of Ginsparg-Wilson fermion,''
Phys.\ Lett.\ B {\bf 445}, 371 (1999)
[arXiv:hep-lat/9809013].
%%CITATION = HEP-LAT 9809013;%%

%\cite{Neuberger:1999pz}
\bibitem{Neuberger:1999pz}
H.~Neuberger,
%``Bounds on the Wilson Dirac operator,''
Phys.\ Rev.\ D {\bf 61}, 085015 (2000)
[arXiv:hep-lat/9911004].
%%CITATION = HEP-LAT 9911004;%%

%\cite{Alvarez-Gaume:1983cs}
\bibitem{Alvarez-Gaume:1983cs}
L.~Alvarez-Gaume and P.~H.~Ginsparg,
%``The Topological Meaning Of Nonabelian Anomalies,''
Nucl. Phys. B {\bf 243}, 449 (1984).
%%CITATION = NUPHA,B243,449;%%

%\cite{Adams:2000yi}
\bibitem{Adams:2000yi}
D.~H.~Adams,
%``Global obstructions to gauge invariance in chiral gauge theory on the
%lattice,''
Nucl. Phys. B {\bf 589}, 633 (2000)
[arXiv:hep-lat/0004015].
%%CITATION = HEP-LAT 0004015;%%

%\cite{Suzuki:2000ii}
\bibitem{Suzuki:2000ii}
H.~Suzuki,
%``Anomaly cancellation condition in lattice gauge theory,''
Nucl.\ Phys.\ B {\bf 585}, 471 (2000)
[arXiv:hep-lat/0002009].
%%CITATION = HEP-LAT 0002009;%%

%\cite{Igarashi:2000zi}
\bibitem{Igarashi:2000zi}
H.~Igarashi, K.~Okuyama and H.~Suzuki,
%``Errata and addenda to 'Anomaly cancellation condition in lattice gauge  theory',''
arXiv:hep-lat/0012018.
%%CITATION = HEP-LAT 0012018;%%

%\cite{Luscher:2000zd}
\bibitem{Luscher:2000zd}
M.~L\"uscher,
%``Lattice regularization of chiral gauge theories to all orders of  perturbation theory,''
JHEP {\bf 0006}, 028 (2000)
[arXiv:hep-lat/0006014].
%%CITATION = HEP-LAT 0006014;%%

%\cite{Kikukawa:2000kd}
\bibitem{Kikukawa:2000kd}
Y.~Kikukawa and Y.~Nakayama,
%``Gauge anomaly cancellations in SU(2)L x U(1)Y electroweak theory on the  lattice,''
Nucl.\ Phys.\ B {\bf 597}, 519 (2001)
[arXiv:hep-lat/0005015].
%%CITATION = HEP-LAT 0005015;%%

%\cite{Kikukawa:2001mw}
\bibitem{Kikukawa:2001mw}
Y.~Kikukawa,
%``Domain wall fermion and chiral gauge theories on the lattice with exact  gauge invariance,''
Phys.\ Rev.\ D {\bf 65}, 074504 (2002)
[arXiv:hep-lat/0105032].
%%CITATION = HEP-LAT 0105032;%%

%\cite{Aoyama:1999hg}
\bibitem{Aoyama:1999hg}
T.~Aoyama and Y.~Kikukawa,
%``A lattice implementation of the eta-invariant and effective action for  chiral fermions on the lattice,''
arXiv:hep-lat/9905003.
%%CITATION = HEP-LAT 9905003;%%

%\cite{Igarashi:2002zz}
\bibitem{Igarashi:2002zz}
H.~Igarashi, K.~Okuyama and H.~Suzuki,
%``More about the axial anomaly on the lattice,''
Nucl.\ Phys.\ B {\bf 644}, 383 (2002)
[arXiv:hep-lat/0206003].
%%CITATION = HEP-LAT 0206003;%%

%\cite{Horvath:1998cm}
\bibitem{Horvath:1998cm}
I.~Horvath,
%``Ginsparg-Wilson relation and ultralocality,''
Phys.\ Rev.\ Lett.\  {\bf 81}, 4063 (1998)
[arXiv:hep-lat/9808002].
%%CITATION = HEP-LAT 9808002;%%

%\cite{Horvath:1999bk}
\bibitem{Horvath:1999bk}
I.~Horvath,
%``Ginsparg-Wilson-Luescher symmetry and ultralocality,''
Phys.\ Rev.\ D {\bf 60}, 034510 (1999)
[arXiv:hep-lat/9901014].
%%CITATION = HEP-LAT 9901014;%%


%\cite{Fujiwara:2000wn}
\bibitem{Fujiwara:2000wn}
T.~Fujiwara, H.~Suzuki and K.~Wu,
%``Topological charge of lattice Abelian gauge theory,''
Prog.\ Theor.\ Phys.\  {\bf 105}, 789 (2001)
[arXiv:hep-lat/0001029].
%%CITATION = HEP-LAT 0001029;%%


\end{thebibliography}
\end{document}